\documentclass[a4paper,fleqn]{cas-dc}

\usepackage{graphicx}
\usepackage[authoryear,round]{natbib}
\usepackage{caption}
\usepackage{accents}

\def\tsc#1{\csdef{#1}{\textsc{\lowercase{#1}}\xspace}}
\tsc{WGM}
\tsc{QE}
\tsc{EP}
\tsc{PMS}
\tsc{BEC}
\tsc{DE}
\begin{document}
\let\WriteBookmarks\relax
\def\floatpagepagefraction{1}
\def\textpagefraction{.001}
\shorttitle{Data-driven enhancement of the attached eddy model for wall turbulence}
\shortauthors{R. Deshpande et~al.}

\title [mode = title]{Data-driven enhancement of coherent structure-based models for predicting instantaneous wall turbulence}

\tnotetext[1]{This work was supported by the Australian Research Council.}

\author[1]{Rahul Deshpande}[orcid=0000-0003-2777-2919]
\cormark[1]
\ead{raadeshpande@gmail.com}

\address[1]{Department of Mechanical Engineering, University of Melbourne, Parkville, Victoria 3010, Australia}
\address[2]{School of Mechanical and Manufacturing Engineering, University of New South Wales, Sydney, NSW 2052, Australia}
\address[3]{Combustion Research Facility, Sandia National Laboratories, Livermore, California 94550, USA}

\author[2]{{Charitha M.} {de Silva}}

\author[3]{Myoungkyu Lee}

\author[1]{Jason P. Monty}

\author[1]{Ivan Marusic}

\cortext[cor1]{Corresponding author}

\begin{abstract}
Predictions of the spatial representation of instantaneous wall-bounded flows, via coherent structure-based models, are highly sensitive to the geometry of the representative structures employed by them.
In this study, we propose a methodology to extract the three-dimensional (3-D) geometry of the statistically significant eddies from multi-point wall-turbulence datasets, for direct implementation into these models to improve their predictions.
The methodology is employed here for reconstructing a 3-D statistical picture of the inertial wall coherent turbulence for all canonical wall-bounded flows, across a decade of friction Reynolds number ($Re_{\tau}$).
These structures are responsible for the $Re_{\tau}$-dependence of the skin-friction drag and also facilitate the inner-outer interactions, making them key targets of structure-based models.
The empirical analysis brings out the geometric self-similarity of the large-scale wall-coherent motions and also suggests the hairpin packet as the representative flow structure for all wall-bounded flows, thereby aligning with the framework on which the attached eddy model (AEM) is based.
The same framework is extended here to also model the very-large-scaled motions, with a consideration of their differences in internal versus external flows.
Implementation of the empirically-obtained geometric scalings for these large structures into the AEM is shown to enhance the instantaneous flow predictions for all three velocity components.
Finally, an active flow control system driven by the same geometric scalings is conceptualized, towards favourably altering the influence of the wall coherent motions on the skin-friction drag.
\end{abstract}


\maketitle

\section{Introduction}
\label{intro}

Owing to its highly chaotic and random nature, an accurate prediction of the instantaneous velocity in a turbulent flow remains the most challenging demand from any turbulence model.
Despite its inherent complexities, turbulence flow modelling has remained an active area of research for over half a century \citep{brunton2020}, given the numerous incentives on offer.
In case of wall-bounded flows, for example, the ability to predict/replicate the instantaneous flow phenomena can greatly facilitate the design of active flow control techniques \citep{choi1994}, and also aid in enhancing future high-fidelity numerical simulations, by providing realistic inflow boundary conditions and improving computational efficiency \citep{subbareddy2006,wu2017}.
One of the popular approaches of modelling wall-turbulence, amongst many, has been by focusing on the recurring and statistically significant `coherent' motions \citep{robinson1991,choi1994} omnipresent in these flows.
These motions play a crucial role in both the kinematics as well as dynamics of wall-bounded flows \citep{robinson1991,jimenez2018,mklee2019} and have been directly associated with the behaviour of several flow statistics -- both averaged \citep{ganapathisubramani2005,hutchins2007,heisel2020} and instantaneous \citep{adrian2000,charitha2016umz,charitha2018}.
Hence, several studies proposing coherent structure-based models can be found in the literature \citep{theodorsen1952,grant1958,willmarth1967,townsend1976,davidson2009}, amongst which the attached eddy model (AEM; \citet{perry1982}, \citet{marusic2019}) based on Townsend's attached eddy hypothesis \citep{townsend1976}, is one of the most cited.
Here, the words `structures', `motions' and `eddies' are used interchangeably and essentially conform to the definition of a coherent motion given by \citet{robinson1991}.

According to Townsend's attached eddy hypothesis, an inviscid asymptotically high friction Reynolds number ($Re_{\tau}$ $=$ $\frac{{U_{\tau}}{\delta}}{\nu}$) canonical wall-bounded flow can be modelled via a hierarchy of geometrically self-similar attached eddies, having population densities inversely proportional to their sizes, which vary over ${\mathcal{O}}$($z$) -- ${\mathcal{O}}$($\delta$).
The term `attached' here refers to any coherent motion whose geometric extent scales with its distance from the wall, given by $z$, while the associated velocity fluctuations scale with the friction velocity $U_{\tau}$, but doesn't necessarily imply its velocity signatures physically extending down to the wall \citep{marusic2019}.
Further, $\delta$ here corresponds to the outer scale of the respective canonical flow geometry, which is the boundary layer thickness for a zero-pressure gradient turbulent boundary layer (ZPG TBL), while it is the pipe radius and channel half-height for internal flows.
Increased access to high $Re_{\tau}$ wall-turbulence datasets, over the past two decades, has confirmed that the kinetic energy production and transfer within these flows is mostly influenced by the inertial (inviscid) motions predominating in the inertial region \citep{marusic2010high,mklee2019,yhwang2020}, making AEM ideally suited for modelling these flows.
These characteristics also make the inertial motions a key target of  active flow control schemes \citep{abbassi2017}.
Studies in the past have already demonstrated that the AEM is able to predict ensemble/time-averaged statistics, such as the Reynolds stresses \citep{marusic1995,baidya2014}, higher order moments \citep{woodcock2015,charitha2016}, two-point correlations \citep{marusic2001} and the energy spectra \citep{baidya2017,deshpande2020,chandran2020} {across the inertial layer (2.6$\sqrt{Re_{\tau}}$ $\lesssim$ $z^+$ $\lesssim$ 0.15$Re_{\tau}$; \citet{klewicki2009}) in high $Re_{\tau}$ canonical flows.}
While there have been a few studies focused on predicting the instantaneous flow phenomena in both the inertial and the wake region \citep{charitha2016umz,charitha2018,eich2020}, success has mostly been limited to the wake region, predominated solely by the largest hierarchy of eddies ($\sim$ $\mathcal{O}$($\delta$)). 

The primary challenge associated with accurately predicting the representative instantaneous flow in the inertial region, is its high sensitivity to the geometry of the coexisting motions, given that the region is populated by a host of statistically significant eddies, with sizes varying from ${\mathcal{O}}$($z$) to ${\mathcal{O}}$($\delta$) \citep{mklee2015,deshpande2020,yoon2020}.
These eddies are highly three-dimensional (3-D) in geometry, and hence vary considerably in terms of their spatial coherence, which also needs to be accounted for while attempting to replicate the instantaneous flow.
To this end, there are several studies in recent literature \citep{yoon2020,charitha2020,chandran2020} which have envisioned and subsequently characterized the inertial region as a superposition of two types of inertial motions: (i) motions with their velocity signatures (i.e. coherence) extending all the way down to the wall (i.e. wall coherent or WC motions) and (ii) those with velocity signatures not extending to the wall (wall incoherent; WI).
Interestingly, both these eddy types (WC and WI) have been shown by the same set of studies to depict characteristics consistent with Townsend’s attached eddies, although with certain caveats; for example, apart from the hierarchy of self-similar eddies, the WC inertial motions also comprise $\delta$-scaled superstructures or very-large-scale motions (VLSMs), which do not conform to Townsend's attached eddies \citep{deshpande2020,yoon2020}.
These results, thus, showcase the prospect of predicting the instantaneous wall-bounded flow by using the AEM framework to individually model the WC and WI motions in the inertial region.

In the present study, we take the first step towards achieving this by establishing an AEM-based methodology to model the inertial WC motions in all three canonical wall-bounded flows. 
Specifically, to improve its predictive capability, the geometry of the attached eddies is based on estimates extracted directly from published experimental and numerical datasets.
It is worth noting here that although the inertial WC motions are only a subset of the full flow, these motions are responsible for the $Re_{\tau}$-dependence of the wall shear stress fluctuations (and hence, the skin-friction drag characteristics) in any wall-bounded flow \citep{orlu2011,deck2014,giovanetti2016,abbassi2017,smits2021}.
Further to that, the WC motions are also known to superimpose onto and modulate the near-wall cycle \citep{hutchins2007,mathis2009}, which has a significant impact on the skin friction drag. 
This makes the present modelling effort relevant for both fundamental as well as applied research.
We begin by first reviewing the current state of the AEM ($\S$\ref{current_AEM}) and its limitations ($\S$\ref{limit_AEM}), followed by proposal of the new methodology ($\S$\ref{3d_stats}), which provides a data-driven basis to improve the AEM predictions.
Throughout this paper, we use the coordinate system $x$, $y$ and $z$ to refer to the streamwise, spanwise and wall-normal directions respectively, with $u$, $v$ and $w$ denoting the corresponding fluctuating velocity components.
${\langle}{\rangle}$ and capitalization indicates averaged quantities while the superscript $+$ refers to normalization in viscous units (eg., $u^+$ $=$ $u$/$U_{\tau}$ and $z^+$ $=$ $z{U_{\tau}}/{\nu}$, where $\nu$ is the kinematic viscosity).

\subsection{Current state of the AEM}
\label{current_AEM}

\begin{figure*}
	\centering
		\includegraphics[width=0.9\textwidth]{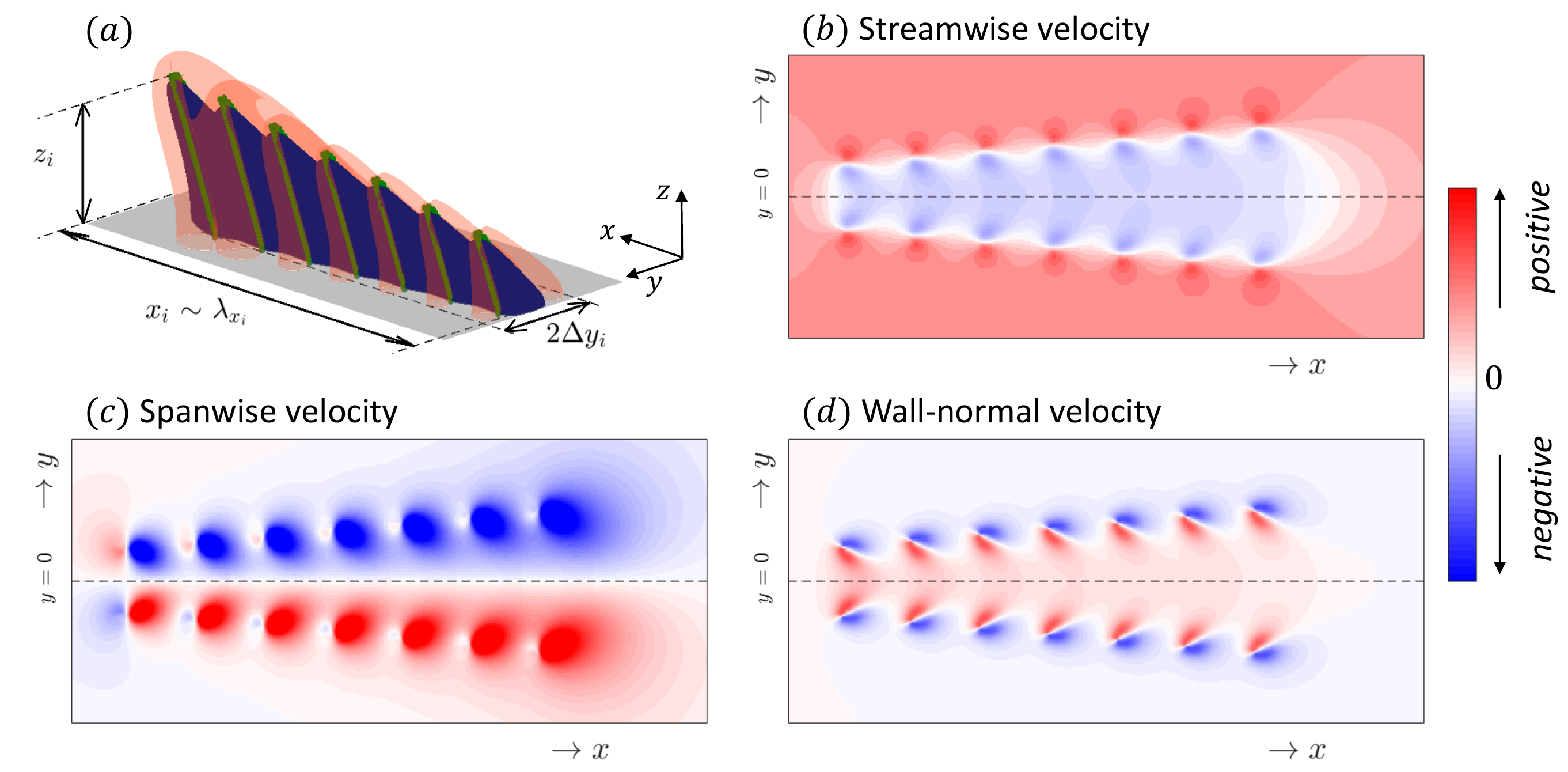}
		\caption{(a) Schematic depicting the representative $\Lambda$-eddy packet used in the AEM \citep{charitha2016,eich2020}, with the blue and red iso-contours respectively denoting the negative and positive streamwise velocity induced from the $\Lambda$-shaped vortex rods (indicated in green). (b) Streamwise, (c) spanwise and (d) wall-normal velocity induced by the $\Lambda$-eddy packet in the reference wall-parallel plane at $z$ = 0.01$z_{i}$ (highlighted in grey) shown in (a). Labels in (a) are used to refer to the 3-D geometry of the packet, with ${\Delta}y_{i}$ representing the spanwise half-width, owing to the symmetry of the $u$-distribution about the $y$ = 0 plane. This figure has been adapted from \citet{charitha2016}. Mean flow direction is along $x$.}
		\label{fig1} 
	%\end{center}
\end{figure*}

In the original AEM conceptualized by \citet{perry1982}, a single hairpin or a simple arch-shaped ($\Lambda$) eddy, inclined forwards with respect to the flow direction ($x$) at $\sim$45$^{\circ}$ \citep{head1981,deshpande2019}, was considered as the representative coherent structure/eddy.
This shape was inspired from the seminal flow visualization studies of \citet{head1981} in a low $Re_{\tau}$ TBL.
The simplest version of this eddy is essentially made up of two vortex rods, arranged in a $\Lambda$-shape, with each rod containing a Gaussian distribution of vorticity about its core. 
The corresponding velocity field for the eddy can be obtained by performing the Biot-Savart calculations (schematically depicted in figures 2, 6 and 7 of \citet{perry1982}).
Over time, the representative eddy shape has evolved as more detailed quantitative measurements, revealing the structure of wall-bounded flows were reported, such as the seminal PIV measurements of \citet{adrian2000}.
While this study highlighted the presence of $\Lambda$-type eddies in the TBL, it was found that the eddies are organized in the form of a packet in fully turbulent flows.
The use of a $\Lambda$-eddy packet (figure \ref{fig1}(a)), instead of a single eddy, as the representative attached eddy was subsequently incorporated into the AEM by \citet{marusic2001}, and this change was shown to further improve statistical predictions.
Figures \ref{fig1}(b-d) depict the near-wall ($z$ = 0.01$z_{i}$) planar flow field of the three velocity components associated with a single packet-like eddy, which is simply a superposition of the velocity fields obtained from multiple individual $\Lambda$-eddies.
In case of the AEM, it should be noted that $w$ = 0 $\neq$ $u$, $v$ at $z$ = 0 due to the impermeability condition being the only condition imposed at the wall, which is achieved by using packet structures with image packet pairs in the plane of the wall.

To model the inertially dominated (i.e. outer) region, the flow is simply represented by the superposition of multiple such $\Lambda$-eddy packets (referred as hierarchies), of varying sizes and population density, randomly distributed in the flow domain.
Following Townsend's hypothesis, the geometry of these hierarchies is considered to vary self-similarly, a claim which has received both empirical \citep{baars2017,deshpande2020,yoon2020,hwang2020} and theoretical \citep{perry1986,mckeon2019} support in the literature.
The recent study by \citet{eich2020} is the latest published AEM flow-field configuration (henceforth referred simply as AEM) used to replicate the instantaneous flow-field for a ZPG TBL at $Re_{\tau}$ $\approx$ 3200.
In this study, the major axis of the $\Lambda$-eddy packet was oriented at various angles (with respect to $x$) along the $x$-$y$ plane, to incorporate the `meandering' characteristics of the eddies noted in experiments \citep{hutchins2007,kevin2019}.
To summarize, while the meandering features and the eddy inclination angles have been chosen based on the empirical estimates \citep{deshpande2019,kevin2019}, the aspect ratios governing the representative eddy geometry (marked in figure \ref{fig1}(a)) have never had an empirical basis to date, which forms the motivation for the present study.
The eddy geometry plays a major role in the visual appearance of the instantaneous flow field generated by the AEM, as will be highlighted in the next section ($\S$\ref{limit_AEM}).

The present study, hence, follows in line with previous studies \citep{charitha2016,eich2020} with the aim of enhancing the AEM framework, in order to improve spatial representation of a wall-bounded flow from the model.
We note that the typical shape of the representative eddy (of a $\Lambda$-eddy packet) has been kept the same with the intention to preserve its low-order complexity, and given its past success in replicating turbulent wall flow statistics \citep{woodcock2015,baidya2014,baidya2017,deshpande2020,chandran2020}.
Further, the present shape is also well suited to model the wall-coherent subset of the wall-bounded flows, given that all three velocity components generated from the eddy extend down to the near-wall region, (figures \ref{fig1}(b-d)). 

\begin{figure*}
	\begin{center}
		\includegraphics[width=0.85\textwidth]{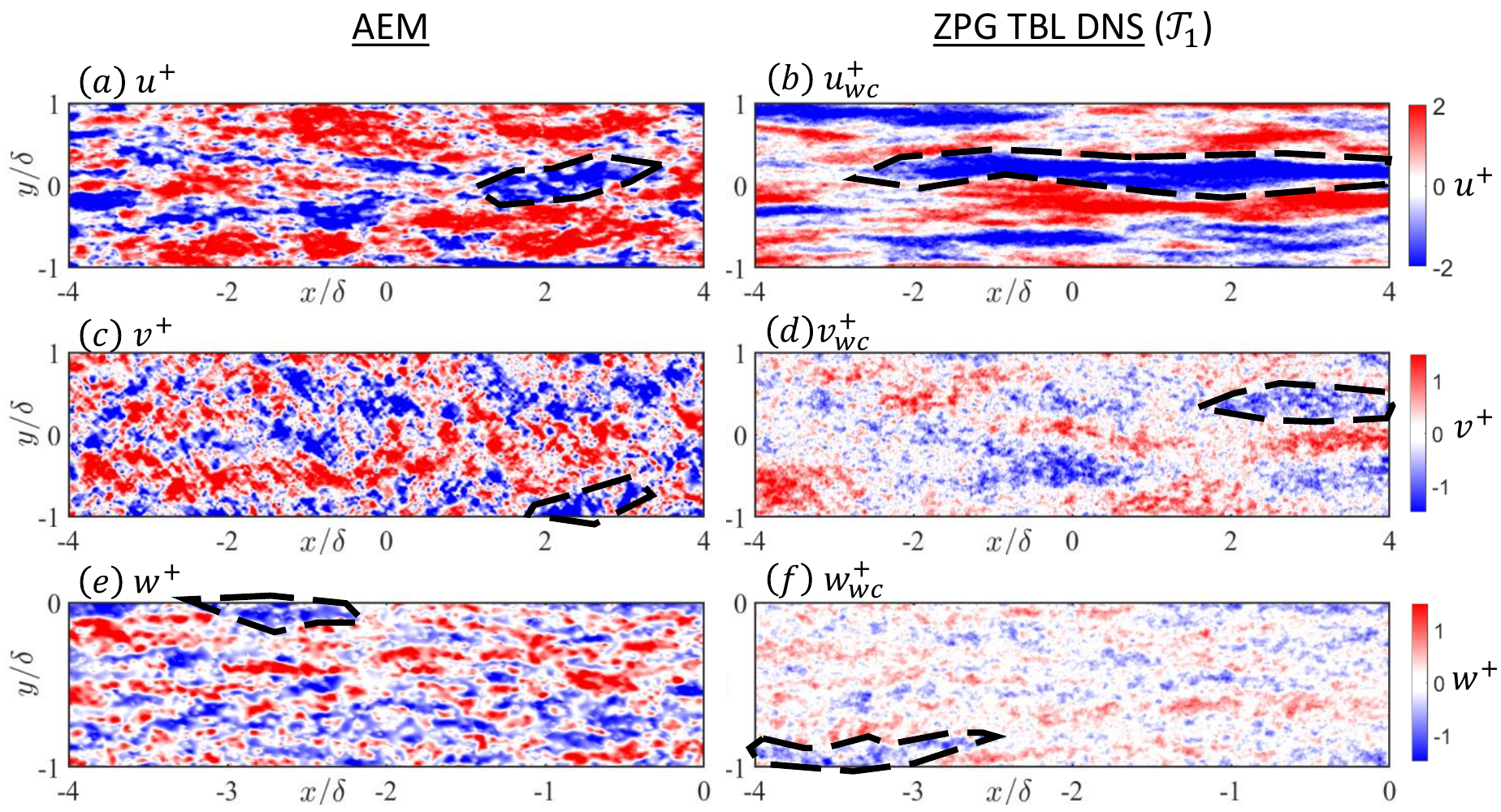}
		\caption{{Instantaneous (a,b) streamwise, (c,d) spanwise and (e,f) wall-normal velocity fluctuations on a wall-parallel plane at the lower bound of the inertial-region ($z^+$ $\approx$ 2.6$\sqrt{Re_{\tau}}$). Data in (a,c,e) corresponds to the AEM of \citet{eich2020} and has been plotted at $z^+$ $\approx$ 160, while that in (b,d,e) corresponds to the ZPG TBL DNS of \citet{sillero2014} comprising solely of wall-coherent motions (represented by subscript wc). The DNS data has been plotted at $z^+$ $\approx$ 100. Black dashed lines are used to highlight the largest spatial features of low momentum for a qualitative comparison. Note the difference in axis limits between (a-d) and (e-f).}}
		\label{fig2} 
	\end{center}
\end{figure*}

\section{Motivation for a data-driven AEM}
\label{limit_AEM}

Here, we compare the instantaneous flow fields generated by the AEM \citep{eich2020} with the corresponding flow fields from the direct numerical simulation (DNS) of a ZPG TBL \citep{sillero2014} to motivate a data-driven definition of the representative eddy geometry used in the former.
Given that the AEM simulates purely the WC portion of the TBL, a logical way to go about this would be by comparing it with the subset of the full DNS fields comprising solely the WC motions.
Figure \ref{fig2} presents this comparison between the instantaneous velocity fluctuations for all three components, considered along the wall-parallel plane at the lower bound of the inertial region.
It brings out significant differences in the geometry of the coherent motions, with the velocity features in the DNS fields substantially longer than those noted in the AEM fields.
A plausible reason behind this difference may be the geometry of a $\Lambda$-eddy packet not defined based on empirical estimates, which we aim to facilitate in the present study.
Here, we propose to obtain the geometric estimates by reconstructing the 3-D statistical picture of the wall-coherent turbulence from published datasets ($\S$\ref{3d_stats}).
Another noteworthy limitation in the AEM is that all the coherent structures considered in the current model correspond to the hierarchy of self-similar eddies, following Townsend's hypothesis.
A true WC flow field, however, cannot be replicated without also considering the $\delta$-scaled superstructures or VLSMs \citep{deshpande2020,yoon2020}, which we also intend to include in the model.
Interestingly, these motions can also be modelled via the same representative $\Lambda$-eddy packet shown in figure \ref{fig1}(a), as demonstrated recently by \citet{chandran2020} for modelling the 2-D spectra.

Readers may note here that the DNS flow fields in figures \ref{fig2}(b,d,f), which comprise solely the WC motions, have been computed using the publicly available full (WC+WI) fields via a spectral linear stochastic estimation (SLSE) based decomposition technique.
Interested readers may refer to appendix \ref{appendix} for more details on how the WC fields were estimated, and also see figure \ref{app_fig1}, where the full flow field instants corresponding to the WC fields in figure \ref{fig2} have been plotted.
Differences between the full (figure \ref{app_fig1}) and the WC (figure \ref{fig2}) flow fields are substantial, especially for the lateral velocity components, with large spatial coherence observed in case of the latter as compared to the former.
Further, the oblique features otherwise apparent in the full spanwise velocity field in the log-region (figure \ref{app_fig1}(b); \citet{sillero2014},\citet{charitha2018}), cannot be noted in its WC subset (figure \ref{fig2}(d)).
These differences underscore the importance of comparing the AEM fields with solely the WC motions, which the representative eddy models.

\begin{table*}
\setlength{\tabcolsep}{2pt}
\caption{Table summarizing the details of five previously published datasets comprising synchronized measurements of $u$-fluctuations by various near-wall fixed probes (placed at $x_{w}$ = 0,$y_{w}$,$z_{w}$) and a traversing probe ($x$ = 0,$y$,$z$), in turn separated by relative spanwise (${\Delta}y$ $=$ ${\mid{y}\;{-}\;{y_w}\mid}$) and wall-normal (${z}\;{-}\;{z_w}$ $\approx$ $z$) offsets.
In case of dataset ${\mathcal{P}}_{1}$, the hot-film attached to the wall measures the instantaneous skin-friction velocity, ${\vec{U}}_{\tau}$ $=$ $\sqrt{{\nu}({{\partial}{\vec{U}}}/{{\partial}z})}$.
The probe arrangements associated with each of these datasets have been schematically depicted in figures \ref{fig3} and \ref{fig4}, with probes in blue and red respectively denoting the traversing and near-wall fixed probes.}
\begin{tabular}{ lccccc }
\toprule
\bf{\underline{Dataset:}} & & & & & \vspace{1.5mm}\\[0.1cm]
$\bullet$ Label & ${\mathcal{T}}_{1}$ & ${\mathcal{C}}_{1}$ & ${\mathcal{T}}_{2}$ & ${\mathcal{T}}_{3}$ & ${\mathcal{P}}_{1}$ \\
$\bullet$ Flow type & ZPG TBL & Channel & ZPG TBL & ZPG TBL & Pipe\\
$\bullet$ Facility & DNS (raw) & DNS (raw) & HRNBLWT & HRNBLWT & CICLoPE\\
$\bullet$ Reference & \citet{sillero2014} & \citet{mklee2015} & \citet{baars2017} & \citet{deshpande2020} & \citet{baidya2019}\\
$\bullet$ $Re_{\tau}$ ${\approx}$ & 2000 & 5200 & 14000 & 14000 & 40000 \vspace{1.0mm}\\[0.1cm]

\bf{\underline{Near-wall probes:}} & & & & & \vspace{2mm}\\[0.1cm]
$\bullet$ Probe type & Grid point & Grid point & Hot-wire & Hot-wire & Hot-film \\
$\bullet$ Number & 38 & 40 & 1 & 1 & 20 \\
$\bullet$ $z^{+}_{w}$ $\approx$ & 15 & 15 & 4.33 & 15 & Wall\\
$\bullet$ Spanwise/azimuthal & 0 $\le$ ${y_{w}}/{\delta}$ $\le$ 1 & 0 $\le$ ${y_{w}}/{\delta}$ $\le$ 1 & $y_{w}$ $\approx$ 0 & $y_{w}$ $\approx$ 0 & 0 $\le$ ${y_{w}}/{\delta}$ $\le$ 1\\
arrangement of probes & (log-spacing) & (log-spacing) & & & (log-spacing) \vspace{1.0mm}\\[0.1cm]

\bf{\underline{Traversing probe:}} & & & & & \vspace{2mm}\\[0.1cm]
$\bullet$ Probe type & Grid point & Grid point & Hot-wire & Hot-wire & Hot-wire \\
$\bullet$ Traversing direction & $z$ & $z$ & $z$ & $y$ & $z$\\
$\bullet$ Traversing range & 15 $<$ $z^{+}$ $\le$ $Re_{\tau}$ & 15 $<$ $z^{+}$ $\le$ $Re_{\tau}$ & 4.33 $<$ $z^{+}$ $\le$ $Re_{\tau}$ & 0 $\le$ $y/{\delta}$ $\le$ $1$ & 0 $<$ $z^{+}$ $\le$ $Re_{\tau}$\\
 & (log-spacing) & (log-spacing) & (log-spacing) & (log-spacing) & (log-spacing) \\
$\bullet$ Position along non- & & & & & \\
traversing directions & $y$ = 0 & $y$ = 0 & $y$ = 0 & $z$ $\sim$ log-region & $y$ = 0 \\
\bottomrule
\end{tabular}
\label{tab_exp}
\end{table*} 

\section{3-D statistical picture of the wall-coherent turbulence}
\label{3d_stats}

\subsection{Datasets and methodology}

The present study utilizes five previously published multi-point datasets, spanning all three canonical flow geometries. Key parameters of the datasets are summarised in table \ref{tab_exp}.
Dataset ${\mathcal{T}_{1}}$ corresponds to DNS of ZPG TBL, while ${\mathcal{T}_{2}}$ and ${\mathcal{T}_{3}}$ correspond to higher $Re_{\tau}$ experimental data for the same flow geometry.
${\mathcal{C}_{1}}$ and ${\mathcal{P}_{1}}$ are respectively the DNS and experimental datasets for a fully turbulent channel and pipe flow.
The selected datasets present a unique combination of synchronously acquired $u$-fluctuations mapped across 3-D space in the wall-bounded shear flow (refer figures \ref{fig3}(a,b) and \ref{fig4}(a-c)).
${\mathcal{T}_{1}}$, ${\mathcal{C}_{1}}$ and ${\mathcal{P}_{1}}$ each comprise $u$-fluctuations acquired using multiple near-wall fixed probes (placed at $x_w$ $=$ 0,$y_w$,$z_w$), distributed in log spacing along the spanwise/azimuthal directions, in conjunction with those acquired by the probe traversing along the wall-normal direction ($z$).
On the other hand, ${\mathcal{T}_{2}}$ and ${\mathcal{T}_{3}}$ comprise $u$-fluctuations from a single near-wall fixed probe (placed at $x_w$ $=$ 0,$y_w$ $=$ 0,$z_w$), acquired synchronously with those measured farther from the wall by a probe traversing along the $z$- and $y$-directions, respectively.
Considering the fact that the canonical wall-bounded flows are statistically homogeneous along the span, the cumulative 3-D space mapped by combining the datasets ${\mathcal{T}_{2}}$ and ${\mathcal{T}_{3}}$ would be equivalent to that mapped in each of the individual datasets, ${\mathcal{T}_{1}}$, ${\mathcal{C}_{1}}$ and ${\mathcal{P}_{1}}$.
In case of ${\mathcal{P}}_{1}$, it should be noted that the near-wall probe is essentially a hot-film sensor attached to the wall for measuring the instantaneous skin-friction velocity ${\vec{U}}_{\tau}$ (= ${U}_{\tau}$ + ${u_{\tau}}$), which is associated with the near-wall instantaneous streamwise velocity (${\vec{U}}$) following ${\vec{U}}_{\tau}$ = $\sqrt{{\nu}({{\partial}{\vec{U}}}/{{\partial}z})}$.
Interested readers may refer to the individual references for more specific details regarding the measurements/simulations.

We now move on to proposing the methodology used to reconstruct the 3-D statistical picture of the wall-coherent turbulence, from which the geometric estimates for the representative eddy (of the AEM) would be extracted.
There have been several studies in the past \citep{ganapathisubramani2005,hutchins2007,lee2011} which have used multi-point datasets to estimate the geometry of the coherent motions, by computing mostly space-time cross-correlations.
These cross-correlations, however, represent cumulative contributions from eddies of various length scales at a specific spatial offset (say streamwise offset, ${\Delta}x$). 
Here, since we are interested in estimating the 3-D geometry of individual hierarchies/length scales to be incorporated in the AEM, we compute the cross-correlations in the spectral domain to get a scale-specific estimate of the coherent structure geometry (i.e. as a function of the streamwise wavelength, ${\lambda}_{x}$). 
Given our focus is on modelling the inertial wall-coherent motions coexisting in the outer region, we consider the cross-correlations specifically between $u$-signals in the outer region (at $z$) and those acquired close to the wall (at $z^{+}_{w}$ $\lesssim$ 15), via the cross-correlation spectra ($\Gamma$) defined as \citep{bailey2008,deshpande2020afmc}:
\begin{equation}
\label{eq1}
\begin{aligned}
{{\Gamma}}({z},{{\Delta}{y}},{\lambda_{x}}) &= \frac{ {\text{Re}}[\{ \widetilde{u}(z,y;\lambda_{x}){{\widetilde{u}}^{\ast}}(z_{w},{y_{w}};\lambda_{x}) \}]}{ {\sqrt{ \{ {\mid {\widetilde{u}(z,y;\lambda_{x})} \mid}^{2} \} }}{\sqrt{ \{ {\mid {\widetilde{u}({z_{w}},{y_{w}};\lambda_{x})} \mid}^{2} \} }}} \\ &= \frac{ \text{Re}[ { {{\phi}'_{{u}{u_{w}}}}(z,z_{w},{\Delta}y;\lambda_{x}) } ] }{ {\sqrt{\phi_{{u}{u}}({z},y;\lambda_{x})}}{\sqrt{\phi_{{u_{w}}{u_{w}}}({z_{w}},{y_{w}};\lambda_{x})}} },  
\end{aligned}
\end{equation}
where ${\tilde{u}}$ = ${\mathcal{F}}$($u$) is the Fourier transform of $u$ in either time or $x$ depending on the dataset with ${\lambda}_{x}$ $=$ 2${\pi}/{k_x}$, where $k_x$ is the streamwise wavenumber. 
Further, the curly brackets ($\{ \}$), asterisk ($\ast$) and vertical bars ($\mid \mid$) indicate the ensemble averaging, complex conjugate and modulus, respectively while Re denotes the real component. 
${\phi}'_{{u}{u_{w}}}$ is, thus, the complex-valued 1-D cross-spectrum between $u$-signals recorded at $z$ and $z_{w}$, which quantifies the scale-specific cross-correlation, while $\phi_{{u}{u}}$ and $\phi_{{u_{w}}{u_{w}}}$ respectively are the conventional 1-D energy spectra at ${z}$ and ${z_{w}}$ used for the scale-specific normalization.
Such a definition forces $\Gamma$ to vary between -1 to 1, with the former and latter respectively indicating perfect anti-correlation and correlation for each scale, ${\lambda}_{x}$.
In case of dataset ${\mathcal{P}_{1}}$, notably, $\Gamma$ is computed by substituting the skin-friction velocity fluctuations ($u_{\tau}$) measured by the wall-based hot-films, in place of $u$($z_w$).

As can be noted from its definition in equation (\ref{eq1}), $\Gamma$ is a function of the spatial offsets in all three directions, which is facilitated by computing the cross-correlations between $u$-signals recorded at various relative spanwise offsets (${\Delta}y$ $=$ ${\mid}{y - {y_w}}{\mid}$) and wall-normal offsets ($z$ $\approx$ ${z - {z_w}}$, given $z_{w}$ $\ll$ $z$).
The offset along $x$ is obtained in the form of a streamwise wavelength, which is estimated by using the Taylor's hypothesis in case of the experimental datasets (where the local convection velocity, $U_{c}$ $=$ $U(z)$), or obtained directly from the spatial domain for the DNS datasets.
$\Gamma$($z$,${\Delta}y$,${\lambda}_{x}$), thus, can be interpreted as a coherence metric which associates each WC scale/eddy (${\lambda}_{x}$) with its corresponding spanwise (${\Delta}y$) and wall-normal extent ($z$).
This information can be directly used to define the geometry of the representative eddy, for the respective coherent structure-based model, corresponding to the streamwise scale, ${\lambda}_{x}$.
The benefit in case of models such as the AEM, which incorporate vorticity-based structures, is that once the eddy geometry of the $\Lambda$-eddy packet is defined, velocity field for all three components can be simulated without needing to estimate $\Gamma$ for the other (lateral) velocity components.

It is worth noting here that $\Gamma$ being used here is different from the linear coherence spectrum (LCS; ${\gamma}^{2}_{L}$) computed previously by \citet{baars2017} and \citet{baidya2019}, since $\Gamma$ retains solely the real part of the cross-spectrum while ${\gamma}^{2}_{L}$ uses the absolute value of the cross-spectrum.
Here, we prefer $\Gamma$ over ${\gamma}^2_{L}$, to define the eddy geometry, given that the anti-correlated regions ($\Gamma$ $<$ 0) are indicative of the relative placement of low-momentum ($-u$) regions with respect to the high-momentum ($+u$) regions, which otherwise can't be inferred from the LCS (0 $\le$ ${\gamma}^2_{L}$ $\le$ 1).
For example, a recent study \citep{deshpande2020afmc} focusing on the $\Gamma$-distribution over very large spatial extents has confirmed the periodic distribution of $\delta$-scaled $+u$ and $-u$ motions along the spanwise direction. 
Investigating the $\Gamma$ distribution in 3-D space, thus, could provide quantitative basis to the choice of the representative eddy for a coherent structure-based model, as will be discussed ahead in this section.

\subsection{Application to high $Re_{\tau}$ ZPG TBL datasets}

\begin{figure*}
	\begin{center}
		\includegraphics[width=1.0\textwidth]{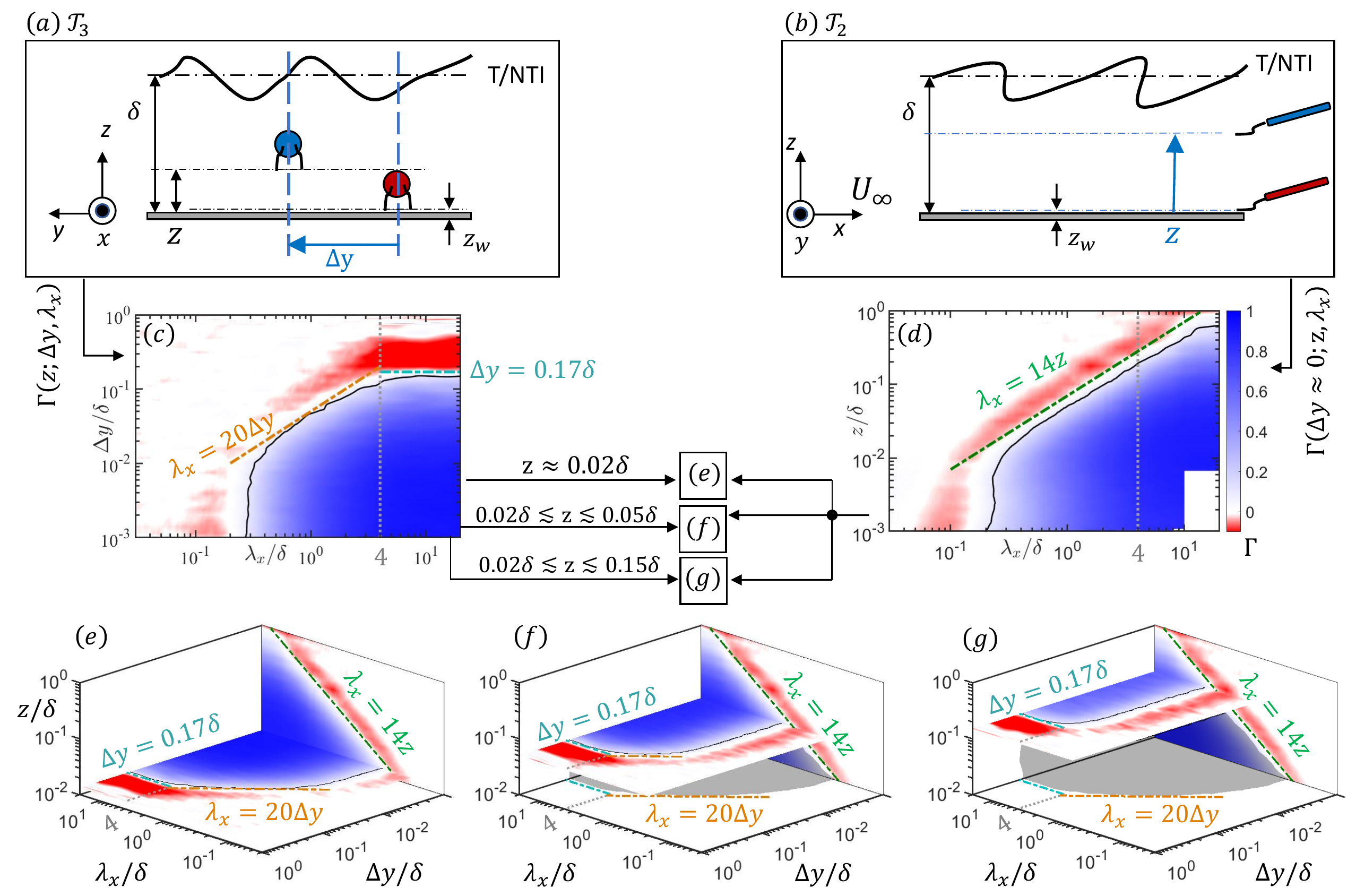}
		\caption{(a,b) Experimental setups corresponding to datasets (a) ${\mathcal{T}}_{3}$ and (b) ${\mathcal{T}}_{2}$. Hotwire sensors marked in blue correspond to traversing probes, which move along their respective traversing directions also marked in blue, while sensors marked in red correspond to the fixed near-wall probes. (c) Cross-correlation spectra computed as a function of ${\Delta}y$ and ${\lambda}_{x}$ (i.e. $\Gamma$(${\Delta}y$,${\lambda}_{x}$)) for $z$ $\approx$ 0.02$\delta$ from the dataset ${\mathcal{T}}_{3}$. (d) $\Gamma$($z$,${\lambda}_{x}$) computed for ${\Delta}y$ $\approx$ 0 from the dataset ${\mathcal{T}}_{2}$. Black contour in (c,d) corresponds to $\Gamma$ = 0.1. (e-g) Reconstruction of the 3-D cross-correlation spectra, $\Gamma$($z$,${\Delta}y$,${\lambda}_{x}$) across the inertial region (0.02$\delta$ $\lesssim$ $z$ $\lesssim$ 0.15$\delta$) by fusing the empirical estimates from datasets ${\mathcal{T}}_{2}$ and ${\mathcal{T}}_{3}$. (e) is obtained by simply combining data in (c) and (d). (f) and (g) are similarly obtained by fusing data in (d) and that estimated from dataset ${\mathcal{T}}_{3}$ over a range of $z$: (f) 0.02$\delta$ $\lesssim$ $z$ $\lesssim$ 0.05$\delta$ and (g) 0.02$\delta$ $\lesssim$ $z$ $\lesssim$ 0.15$\delta$. Black iso-contour in (e-g) also corresponds to $\Gamma$ = 0.1. Dash-dotted lines in green, orange, indigo and grey colours represent the scalings: ${\lambda}_{x}$ = 14$z$, ${\lambda}_{x}$ = 20${\Delta}y$, ${\Delta}y$ = 0.17$\delta$ and ${\lambda}_{x}$ = 4$\delta$, respectively.}
		\label{fig3} 
	\end{center}
\end{figure*}

We begin the empirical analysis by first reconstructing $\Gamma$($z,{{\Delta}y},{{\lambda}_{x}}$) for high $Re_{\tau}$ ZPG TBL using the two experimental datasets, ${\mathcal{T}}_{3}$ and ${\mathcal{T}}_{2}$.
In case of ${\mathcal{T}}_{2}$, the traversing probe is constrained to move vertically above the near-wall fixed probe (i.e. maintaining ${\Delta}y$ $=$ 0), owing to which this dataset yields the statistical picture solely in the streamwise wall-normal plane, given by $\Gamma$(${{\Delta}y}$ $=$ 0$;z,{{\lambda}_{x}}$) in figure \ref{fig3}(d).
This plot brings out the range of energetic WC scales/motions coexisting in the outer region, which correspond to the wavelength range, ${\lambda}_{x}$ $\gtrsim$ 0.1$\delta$.
For a specific energetic length scale, say ${\lambda}_{x}$ $\sim$ 4$\delta$, it indicates the presence of a positively correlated region up to $z$ $\sim$ 0.3$\delta$, followed by a negatively correlated region between 0.3$\delta$ $<$ $z$ $<$ 0.6$\delta$.
{When interpreted physically in terms of the spatial extent of the $u$-velocity distributions along the mid-span of a $\Lambda$-eddy packet of length $\sim$ 4$\delta$ (figure \ref{fig1}(a)), the aforementioned $z$-ranges may be respectively associated with the wall-normal extent of the $-u$ region (represented by +$\Gamma$) and of the $+u$ region (represented by -$\Gamma$).}
Consequently, the height of the $\Lambda$-eddy packet may be nominally defined based on the region where $\Gamma$ $\approx$ 0, which is approximately $z$ $\sim$ 0.3$\delta$.
If we consider this for the entire range ${\lambda}_{x}$ $\gtrsim$ 0.1$\delta$, figure \ref{fig3}(d) indicates that the height ($z$) of the inertially dominated eddies varies self-similarly with respect to their streamwise length scale (${\lambda}_{x}$), which is given by the linear relationship ${\lambda}_{x}$ $=$ 14$z$ (indicated by a dashed-dotted green line fitted to $\Gamma$ $\approx$ 0).
Interestingly, the same linear relationship between ${\lambda}_{x}$ and $z$ was noted by \citet{baars2017} and \citet{baidya2019} for a ZPG TBL and a pipe respectively, although based on ${\gamma}^2_{L}$ as the metric.
{The $\Gamma$-contours, however, can be noted to be deviating from the linear relationship for $z$ $<$ 0.02$\delta$, which corresponds to the nominal lower bound of the log-region ($z/{\delta}$ $\sim$ 2.6$(Re_{\tau})^{-0.5}$) as per \citet{klewicki2009}. 
This deviation suggests that only ${{\lambda}_{x}}/{\delta}$ $>$ 0.3 ($\sim$ 14$\times$2.6$(Re_{\tau})^{-0.5}$) conform to the truly self-similar hierarchy at this flow $Re_{\tau}$.}

The same analysis is next conducted along the wall-parallel plane, at $z$ corresponding to the inertial-region, by using the dataset ${\mathcal{T}}_{3}$.
Figure \ref{fig3}(c) plots $\Gamma$($z$ $=$ 0.02$\delta$;${\Delta}y$,${\lambda}_{x}$) as an example, which corresponds to the 2-D statistical picture at the lower bound of the inertial region.
This picture, however, represents the spanwise coherence only along the $+{\Delta}y$ axis and would be complete by considering a mirror image of the same plot about ${\Delta}y$ = 0, owing to symmetry \citep{hutchins2007}.
Given that the $\Gamma$-distribution in figure \ref{fig3}(c) is of the same nature as in figure \ref{fig3}(d), the physical interpretation discussed for the latter is now extended to the former.
By fitting a dashed-dotted orange line along $\Gamma$ $\approx$ 0, we can interpret from figure \ref{fig3}(c) that the spanwise half-width (${\Delta}y$) of the $\Lambda$-eddy packet also varies self-similarly with respect to ${\lambda}_x$, which is represented by the linear relationship ${\lambda}_{x}$ $=$ 20${\Delta}y$.
Consistent with what was noted from figure \ref{fig3}(d), the self-similar trend is observed for $\lambda_{x}$ $>$ 0.3$\delta$ but is only valid up to $\lambda_{x}$ $\sim$ 4$\delta$, as all the larger length scales are found to have a constant half-width, ${\Delta}y$ $\sim$ 0.17$\delta$.
These large eddies, hence, do not conform to Townsend's hierarchy of attached eddies but, in fact, correspond to the $\delta$-scaled very-large-scales or superstructures noted previously in the ZPG TBL \citep{hutchins2007,lee2011}.
The spanwise widths of these structures, which is found from the present analysis to be 2${\Delta}y$ $\sim$ 0.35$\delta$, is consistent with previous findings based on PIV experiments \citep{dennis2011part2,gao2011}.

Following the analysis on the individual 2-D planes, the plots in figures \ref{fig3}(c,d) can now be stitched together to reconstruct the 3-D statistical picture of the WC motions in the ZPG TBL, as shown in figure \ref{fig3}(e). %, with the information along ${\Delta}y$ limited solely to $z$ $\approx$ 0.01$\delta$.
By estimating the cross-correlation spectra at various $z$: 0.02$\delta$ $\lesssim$ $z$ $\lesssim$ 0.15$\delta$ from dataset ${\mathcal{T}}_{3}$, $\Gamma$($z$,${\Delta}y$,${\lambda}_{x}$) can be reconstructed across the inertial region, as depicted in figures \ref{fig3}(e-g).
These figures, thus, describe the scale-specific geometry of the range of energetic motions coexisting in the inertial region:
the flow at the beginning of the inertial region ($z$ $\sim$ 0.02$\delta$) comprises contributions from the widest hierarchy of self-similar eddies, spanning 0.3$\delta$ $\lesssim$ ${\lambda}_{x}$ $\lesssim$ 4$\delta$, which reduces to 0.8$\delta$ $\lesssim$ ${\lambda}_{x}$ $\lesssim$ 4$\delta$ at $z$ $\approx$ 0.05$\delta$, and finally negligible contributions from the self-similar hierarchy at the upper bound ($z$ $\approx$ 0.15$\delta$).
The contributions from the $\delta$-scaled superstructures (${\lambda}_{x}$ $\gtrsim$ 4$\delta$), on the other hand, is consistently present throughout the inertial region.
The corresponding spanwise and wall-normal extents for each eddy (${\lambda}_{x}$) can also be inferred based on the scalings shown in figure \ref{fig3}(g), and have been summarized as a function of the flow $Re_{\tau}$ below:
\begin{equation}
\label{eq2}
{{\lambda}^{+}_x} {\gtrsim} 4{Re_{\tau}} \; \begin{cases}
                          {z^{+}} = {{\lambda}^{+}_x}/14; {\Delta}y^{+} = 0.17{Re_{\tau}} & \text{if true}\\
                          
                          {z^{+}} = {{\lambda}^{+}_x}/14; {\Delta}y^{+} = {{\lambda}^{+}_x}/20 & \text{otherwise}.
                          \end{cases}       
\end{equation}
This information can be directly used to facilitate definition of the representative eddy geometries in any coherent structure-based model.
It is worth noting here though, $\Gamma$($z$,${\Delta}y$,${\lambda}_{x}$) in figure \ref{fig3}(g) looks consistent with the shape of a $\Lambda$-eddy packet when halved along the span (refer to figure \ref{fig1} or \ref{fig5}), thus providing quantitative support to the choice of the representative eddy shape used in the AEM and making a strong case in favour of implementing equation (\ref{eq2}) into AEM.

\subsection{Extension to various $Re_{\tau}$ and flow geometries}

\begin{figure*}
	\begin{center}
		\includegraphics[width=1.05\textwidth]{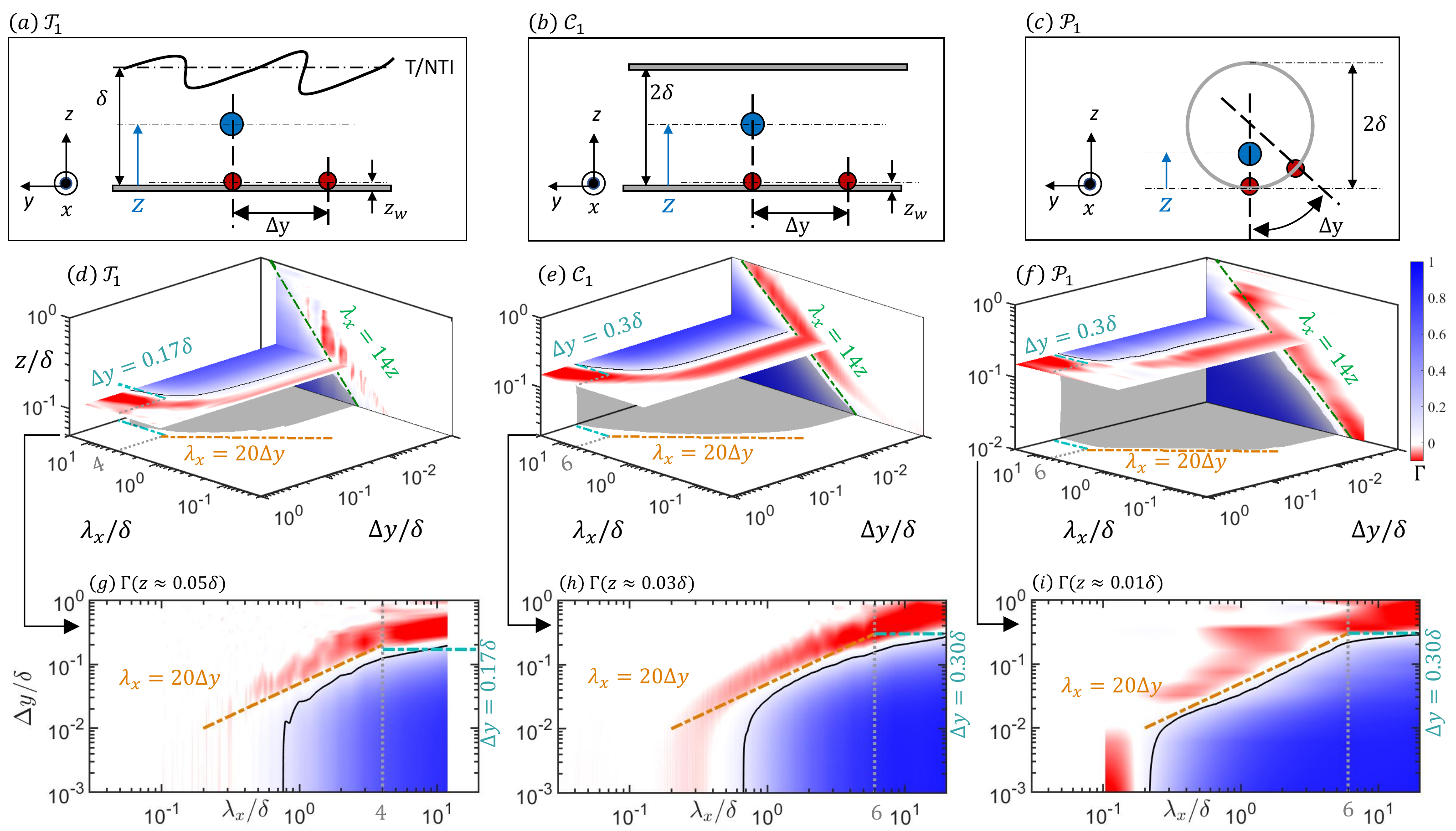}
		\caption{{Reconstruction of the (d-f) 3-D cross-correlation spectra, $\Gamma$($z$,${\Delta}y$,${\lambda}_{x}$) across the inertial region and (g-i) $\Gamma$(${\Delta}y$,${\lambda}_{x}$) at the lower bound of the log-region ($z^+$ $\approx$ 2.6$\sqrt{Re_{\tau}}$) for datasets (d,g) ${\mathcal{T}}_{1}$, (e,h) ${\mathcal{C}}_{1}$ and (f,i) ${\mathcal{P}}_{1}$ by utilizing the $u$-fluctuations associated with the probe placements depicted for the respective flow geometries in (a-c). 
Probes marked in blue correspond to traversing probe which move along their respective traversing directions also marked in blue, while those marked in red correspond to the fixed near-wall probes. Here, only two near-wall probes are shown for representative purposes, with the actual number given in table \ref{tab_exp}. Black iso-contour in (d-f) corresponds to $\Gamma$ = 0.1. Dash-dotted lines in green, orange, indigo and grey colours represent scalings indicated in each figure.}} 
		\label{fig4} 
	\end{center}
\end{figure*}

Having described the methodology to reconstruct ${\Gamma}$($z$,${\Delta}y$,${{\lambda}_{x}}$) and applied it to high $Re_{\tau}$ ZPG TBL data, we now extend the same to datasets at different $Re_{\tau}$ (${\mathcal{T}}_{1}$) and other flow geometries (${\mathcal{C}}_{1}$ and ${\mathcal{P}}_{1}$) to check for the universality of the scalings noted in figure \ref{fig3}.
As discussed previously, each of these datasets are self-sufficient to reconstruct the 3-D statistical picture across the inertial region.
Figure \ref{fig4}(d) plots the same for the low $Re_{\tau}$ ZPG TBL data using similar plotting style as in figure \ref{fig3}(g).
Interestingly, all the scalings noted previously from the high $Re_{\tau}$ experimental data (expressed in equation (\ref{eq2})) are also observed for the low $Re_{\tau}$ case, confirming the $Re_{\tau}$-invariance of these empirical estimates and consequently, their direct utilization in data-driven coherent structure-based models for flows at any $Re_{\tau}$.
Further, the spanwise half-width (${\Delta}y$) is also found to tend towards a constant ($\sim$ 0.17$\delta$) at ${\lambda}_{x}$ $\gtrsim$ 4$\delta$ for the DNS dataset, confirming that the wavelength range estimated for the $\delta$-scaled superstructures is independent of the Taylor's hypothesis assumption.
{It is worth noting here that the low $Re_{\tau}$ for the dataset ${\mathcal{T}}_{1}$ reduces the thickness of the inertial region (0.05$\delta$ $\lesssim$ ${\lambda}_{x}$ $\lesssim$ 0.15$\delta$) and consequently the range of energetic scales (${\lambda}_{x}$) corresponding to the self-similar hierarchy.
This can be confirmed from $\Gamma$($z$ $\approx$ 0.05$\delta$;${\lambda}_{x}$,${\Delta}y$) plotted in figure \ref{fig4}(g), where the contours can be seen to follow the linear relationship only for ${{\lambda}_{x}}/{\delta}$ $>$ 0.8, which is consistent with the $Re_{\tau}$-dependance of the lower limit of the self-similar scaling, as noted previously in the high $Re_{\tau}$ dataset in figure 3(c).}

Figures \ref{fig4}(e,h) and (f,i) respectively depict ${\Gamma}$($z$,${\Delta}y$,${{\lambda}_{x}}$) reconstructed using the high $Re_{\tau}$ channel (${\mathcal{C}}_{1}$) and pipe flow (${\mathcal{P}}_{1}$) datasets.
Remarkably, the same self-similar scalings for the WC motions, noted previously for ZPG TBL, are also noted for the case of internal flows.
Similarly, it is found that the contribution from the self-similar hierarchies becomes statistically insignificant beyond the upper bound of the inertial region, in both the internal flows.
Although not shown here, ${\Gamma}$($z$,${\Delta}y$,${{\lambda}_{x}}$) (similar to figure \ref{fig4}(f)) was also reconstructed for the relatively low $Re_{\tau}$ data ($\approx$ 10000 and 20000) acquired at the same facility by \citet{baidya2019}, which exhibited the same scalings/behaviour.
This universality in the self-similar scaling, across all three canonical flows, is consistent with the observation of \citet{hwang2020} based on low $Re_{\tau}$ ($\approx$ 1000) DNS datasets.
Given that the estimation of the exact linear scalings at low $Re_{\tau}$ may be limited by the narrower range of the self-similar hierarchy, the present study confirms the existence as well as the $Re_{\tau}$-invariance of these self-similar scalings across all three canonical flows.
{Further, given the channel and pipe datasets are at differing $Re_{\tau}$, the $\Gamma$ contours (figure \ref{fig4}(h,i)) also exhibit the $Re_{\tau}$-dependance of the lower-limit of the self-similar range. 
Consistent with the observations noted for the ZPG TBL, we can see that an increase in $Re_{\tau}$ extends the linear scaling up to much smaller scales, where the limit is governed by ${{\lambda}_{x}}/{\delta}$ $\sim$ $(Re_{\tau})^{-0.5}$.}

{In case of the internal flows, however, the ${\lambda}_{x}$ = 20${\Delta}y$ relationship is found to be valid up to a larger length scale ($\lambda_{x}$ $\approx$ 6$\delta$) than the ZPG TBL ($\lambda_{x}$ $\approx$ 4$\delta$), beyond which the constant spanwise half-width gradually plateaus to a constant, ${\Delta}y$ $\approx$ 0.3$\delta$.}
The scalings describing the 3-D spatial extent of the inertial motions, in case of the internal flows, can thus be expressed as a function of $Re_{\tau}$ following:
\begin{equation}
\label{eq3}
{{\lambda}^{+}_x} {\gtrsim} 6{Re_{\tau}} \; \begin{cases}
                          {z^{+}} = {{\lambda}^{+}_x}/14; {\Delta}y^{+} = 0.30{Re_{\tau}} & \text{if true}\\
                          
                          {z^{+}} = {{\lambda}^{+}_x}/14; {\Delta}y^{+} = {{\lambda}^{+}_x}/20 & \text{otherwise}.
                          \end{cases}       
\end{equation}
These trends suggest that the $\delta$-scaled VLSMs (${\lambda}_{x}$ $\gtrsim$ 6$\delta$) conforming to the internal flow geometries are relatively wider than the superstructures in ZPG TBL \citep{monty2009}.
These differences between the internal and external flows have been previously noted by \citet{balakumar2007}, \citet{monty2007} and \citet{lee2013} and were attributed to the `persistent growth' of the energetic motions, beyond the inertial region, in case of internal flows, which is otherwise inhibited by the turbulent/non-turbulent interface (T/NTI) in the ZPG TBL \citep{monty2007,lee2013}. 

\section{Data-driven AEM}
\label{dd-AEM}

\subsection{Incorporating the empirically-obtained scaling laws into the AEM}

\begin{figure}
	\begin{center}
		\includegraphics[width=0.5\textwidth]{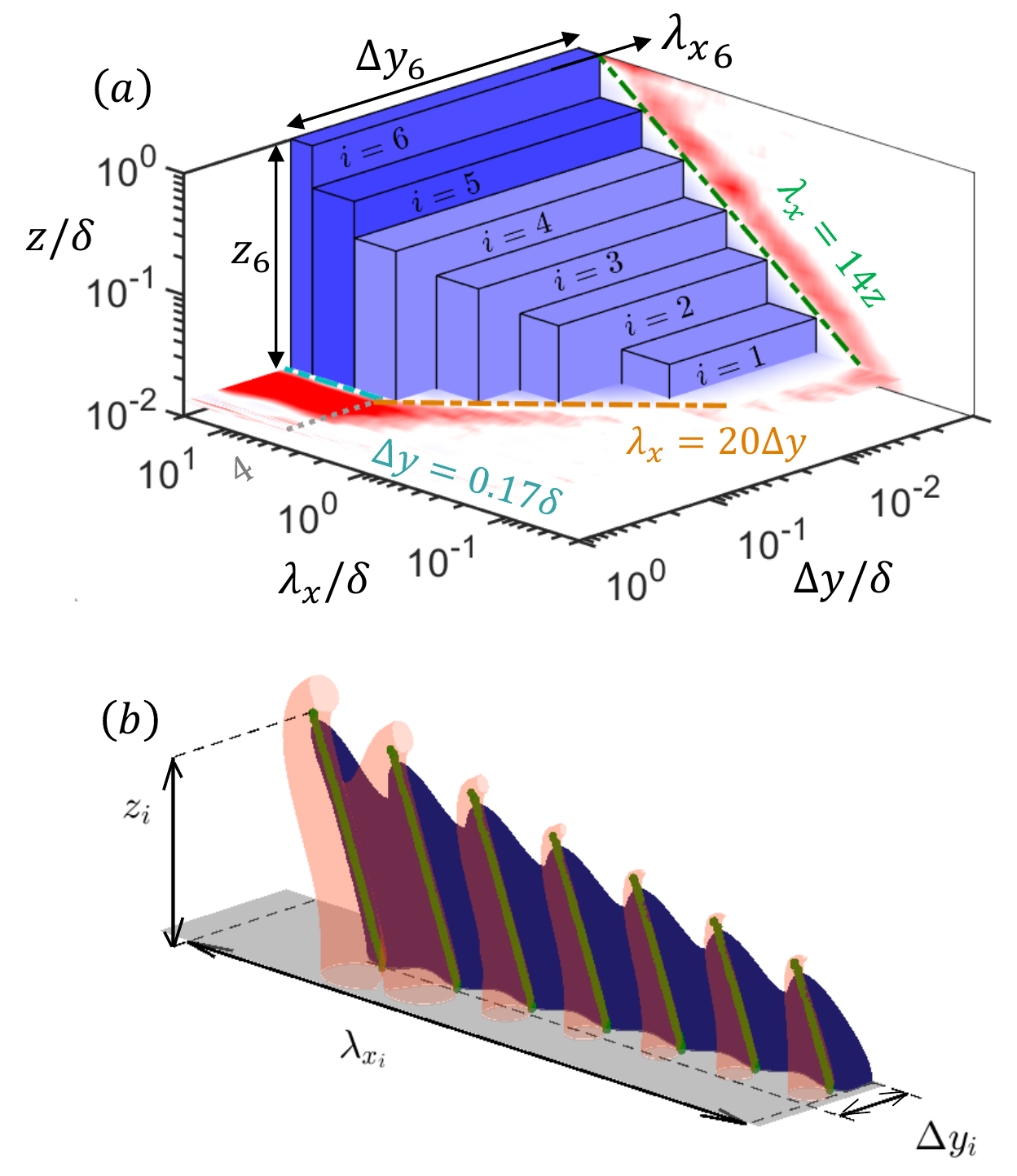}
		\caption{(a) A discretized representation of $\Gamma$($z$,${\Delta}y$,${\lambda}_{x}$) for a ZPG TBL as the superposition of six discrete eddy hierarchies ($i$ = 1 to 6), following the attached eddy model of \citet{perry1982}. Light-shaded boxes represent the self-similar hierarchy while the dark-shaded boxes represent the $\delta$–scaled superstructures. Dash-dotted lines in green, orange, indigo and grey colours represent the same scalings as described in figure \ref{fig3}. Blue and red colour contours plotted along the ${\Delta}y$ = 0 and $z$ $\approx$ 0.01$\delta$ planes are from figure \ref{fig3}(e) and are used simply for reference. (b) The spanwise half of the same $\Lambda$-eddy packet as in figure \ref{fig1}(a), representing the $i^{th}$ hierarchy with its geometry defined in accordance to the discretized plot of $\Gamma$($z$,${\Delta}y$,${\lambda}_{x}$) shown in (a). Table \ref{tab_AEM} discusses the geometry of each of the six eddy hierarchies interpreted from (a).} 
		\label{fig5} 
	\end{center}
\end{figure}

\begin{table*}
\centering
\caption{Comparing the geometry of the various eddy hierarchies in the AEM \citep{eich2020} and dd-AEM. 
Here, SS and NSS respectively denote self-similar and non-self-similar. Other terminologies have been defined in figure \ref{fig5}(b).
$\Lambda$-eddy packets corresponding to the AEM and dd-AEM comprise of 7 and 30 $\Lambda$-eddies respectively, to account for the differing ${\lambda}_{x_{i}}$/$z_{i}$.}
\begin{center}
\begin{tabular}{ccccccccc}
\hline
\\
& & \multicolumn{3}{c}{AEM \citep{eich2020}} & & \multicolumn{3}{c}{dd-AEM (present study)} \\
\cmidrule{3-5} \cmidrule{7-9}
Hierarchy, $i$ & $z_{i}/{\delta}$ & ${\lambda}_{x_{i}}$ & ${\Delta}y_{i}$ & SS/NSS & & ${\lambda}_{x_{i}}$ & ${\Delta}y_{i}$ & SS/NSS\\
\vspace{0.15mm}\\
1 & $z_1$ = 2$^{-5}$ & ${\lambda}_{x_{1}}$ = 3$z_1$ & ${\Delta}y_1$ = ${\lambda}_{x_{1}}$/12 & SS & & ${\lambda}_{x_{1}}$ = 14$z_1$ & ${\Delta}y_1$ = ${\lambda}_{x_{1}}$/20 & SS\\
\vspace{0.15mm}\\
2 & $z_2$ = 2$^{-4}$ & ${\lambda}_{x_{2}}$ = 3$z_2$ & ${\Delta}y_2$ = ${\lambda}_{x_{2}}$/12 & SS & & ${\lambda}_{x_{2}}$ = 14$z_2$ & ${\Delta}y_2$ = ${\lambda}_{x_{2}}$/20 & SS\\
\vspace{0.15mm}\\
3 & $z_3$ = 2$^{-3}$ & ${\lambda}_{x_{3}}$ = 3$z_3$ & ${\Delta}y_3$ = ${\lambda}_{x_{3}}$/12 & SS & & ${\lambda}_{x_{3}}$ = 14$z_3$ & ${\Delta}y_3$ = ${\lambda}_{x_{3}}$/20 & SS\\
\vspace{0.15mm}\\
4 & $z_4$ = 2$^{-2}$ & ${\lambda}_{x_{4}}$ = 3$z_4$ & ${\Delta}y_4$ = ${\lambda}_{x_{4}}$/12 & SS & & ${\lambda}_{x_{4}}$ = 14$z_4$ & ${\Delta}y_4$ = ${\lambda}_{x_{4}}$/20 $\approx$ 0.17$\delta$ & SS\\
\vspace{0.15mm}\\
5 & $z_5$ = 2$^{-1}$ & ${\lambda}_{x_{5}}$ = 3$z_5$ & ${\Delta}y_5$ = ${\lambda}_{x_{5}}$/12 & SS & & ${\lambda}_{x_{5}}$ = 14$z_5$ & ${\Delta}y_5$ $\approx$ 0.17$\delta$ & NSS\\
\vspace{0.15mm}\\
6 & $z_6$ = 2$^{0}$ & ${\lambda}_{x_{6}}$ = 3$z_6$ & ${\Delta}y_6$ = ${\lambda}_{x_{6}}$/12 & SS & & ${\lambda}_{x_{6}}$ = 14$z_6$ & ${\Delta}y_6$ $\approx$ 0.17$\delta$ & NSS\\
\\
\hline
\end{tabular}
\label{tab_AEM}
\end{center}
\end{table*}

With the 3-D statistical picture of the wall-coherent turbulence now characterized for all three canonical flows, we move onto discussing how the empirically obtained scalings can be used in the data-driven AEM (dd-AEM).
In the remainder of this manuscript, we concentrate our efforts on the attached eddy modelling of solely the ZPG TBL flow.
The methodology, however, can also be applied to the internal flows by using the appropriate empirical estimates (given in equation (\ref{eq3})).
Although a real ZPG TBL flow comprises of a continuous distribution of statistically energetic motions over a broadband spectrum (figure \ref{fig3}(g)), for the case of the AEM, we consider a discretized distribution of eddies across $n$ hierarchies for convenience in modelling.
Here, $n$ depends on the $Re_{\tau}$ of the flow being modelled, with higher $Re_{\tau}$ requiring the inclusion of more hierarchies/scales.
{While the wall-normal height of the largest hierarchy, say $z_{n}$ = $\delta$ by definition, we follow the convention adopted by \citet{perry1982} of defining the wall-normal extent of the smallest hierarchy (say $z_{1}$) equal to the nominal lower bound of the outer region, i.e. $z^{+}_{1}$ $\sim$ 100.
\citet{perry1982} further defined the heights of the subsequently larger hierarchies to be double that of the relatively small hierarchy, which can be expressed mathematically as $z^{+}_{i}$ $=$ ${{\delta}^{+}}{2^{i-n}}$, with $i$ being the hierarchy number. 
For $z^{+}_{1}$ $=$ 100, ${\delta}^{+}$ $\sim$ $Re_{\tau}$ $=$ 100($2^{n-1}$), meaning the number of hierarchies $n$ essentially define the $Re_{\tau}$ of the flow.

\begin{figure*}
	\begin{center}
		\includegraphics[width=0.99\textwidth]{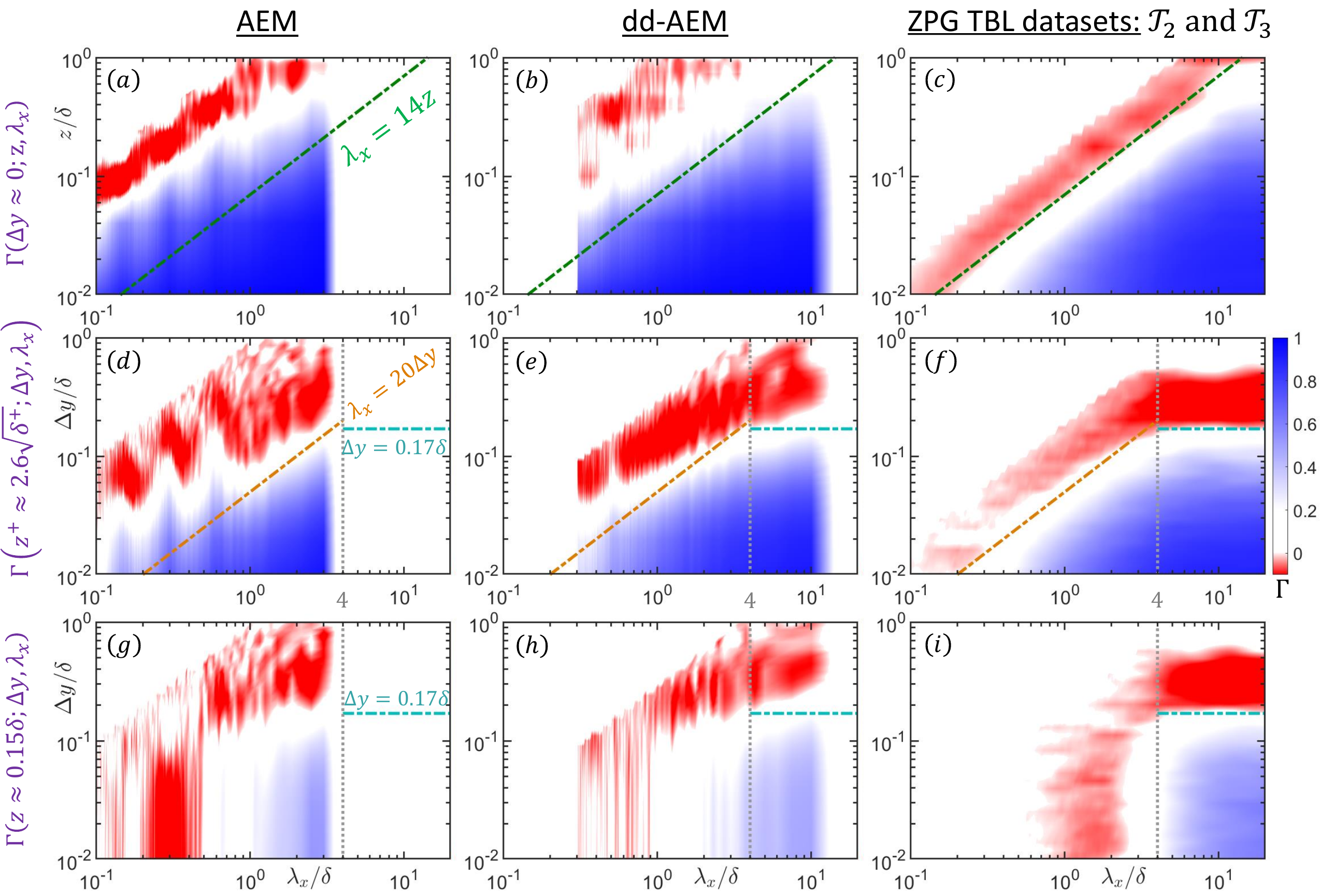}
		\caption{Cross-correlation spectra computed along (a-c) XZ-plane at ${\Delta}y$ $\approx$ 0, $\Gamma$(${\Delta}y$ $\approx$ 0;$z$,${\lambda}_{x}$), (d-f)  XY-plane at the lower bound of the inertial region, $\Gamma$($z^+$ $\approx$ 2.6$\sqrt{Re_{\tau}}$;${\Delta}y$,${\lambda}_{x}$) and (g-i) XY-plane at the upper bound of the inertial region, $\Gamma$($z$ $\approx$ 0.15$\delta$;${\Delta}y$,${\lambda}_{x}$) from the (a,d,g) AEM of \citet{eich2020}, (b,e,h) data driven-AEM and (c,f,i) datasets ${\mathcal{T}}_{2}$ and ${\mathcal{T}}_{3}$. Dash-dotted lines in green, orange, indigo and grey colours represent the same scalings as described in figure \ref{fig3}. The plots in (c,f) represent the same data as in figures \ref{fig3}(d,c) respectively, with the only difference being the change in axis limits to match with the plots from the AEM.} 
		\label{fig6} 
	\end{center}
\end{figure*}

Here, we follow \citet{eich2020} and use $n$ = 6 for the dd-AEM, yielding $Re_{\tau}$ $\approx$ 3200 for the simulations which is also close to the DNS at $Re_{\tau}$ $\approx$ 2000.}
To extract the geometric estimates for the eddy packet (figure \ref{fig5}(b)) corresponding to each of the six hierarchies, figure \ref{fig5}(a) shows the same 3-D statistical picture as in figure \ref{fig3}(g) discretized into six individual blocks.
Here, the $x$-axis location of the $i^{th}$ block represents the streamwise length scale (${\lambda}_{x_{i}}$) of the $i^{th}$ hierarchy, while the corresponding extents along the $y-$ and $z-$axis respectively indicate its spanwise half-width (${\Delta}y_{i}$) and the wall-normal height ($z_{i}$).
It is worth noting here that the eddy packet in figure \ref{fig5}(b), with seven $\Lambda$-eddies, is simply shown for representative purposes and the actual number of $\Lambda$-eddies in the packet may vary depending on ${\lambda}_{x_{i}}$/$z_{i}$ (table \ref{tab_AEM}), such that the inter-eddy spacing is maintained constant for both AEM and dd-AEM.

Table \ref{tab_AEM} records the geometric estimates of all six hierarchies for the dd-AEM, and compares them with those considered in the previous AEM by \citet{eich2020}. 
The size of hierarchies 1 to 4 grows self-similarly in all three directions, in accordance to the empirically obtained scalings, which associates them with Townsend's attached eddies.
Hierarchies 5 and 6, on the other hand, have a constant spanwise half-width (${\Delta}y$ $\sim$ 0.17$\delta$).
The two largest hierarchies, thus, represent the $\delta$-scaled superstructures, which were found to be the sole energetic WC motions coexisting beyond the upper bound of the inertial region (figure \ref{fig3}(g)).
The presence of these $\delta$-scaled eddies, which do not conform to the self-similar hierarchy, is another improvement of the dd-AEM over the AEM, given that all six hierarchies in the latter conformed to the self-similar hierarchy (table \ref{tab_AEM}).

The best way to quantify the impact of the changed eddy aspect ratios (table \ref{tab_AEM}) in the AEM would be to compute the cross-correlation spectra ($\Gamma$) using both AEM and dd-AEM fields and comparing it with the empirical estimate.
This comparison is showcased in figure \ref{fig6}, where $\Gamma$ is considered along the XZ plane (figures \ref{fig6}(a-c)), and that along two wall-parallel planes at the lower (figures \ref{fig6}(d-f)) and upper bound of the inertial region (figures \ref{fig6}(g-i)).
Here, for convenience in interpretation, $\Gamma$ computed from both the AEMs is considered only in the ${\lambda}_{x}$-range encompassing all six hierarchies (table \ref{tab_AEM}).
We can note a significant difference between $\Gamma$ estimated from the AEM and the dd-AEM fields, with the latter following the scaling trends consistent with those obtained empirically.
Based on the comparison along both the wall-parallel planes (figures \ref{fig6}(d-i)), we can expect the dd-AEM to perform much better than the AEM throughout the inertial region.

\begin{figure*}
	\begin{center}
		\includegraphics[width=1.05\textwidth]{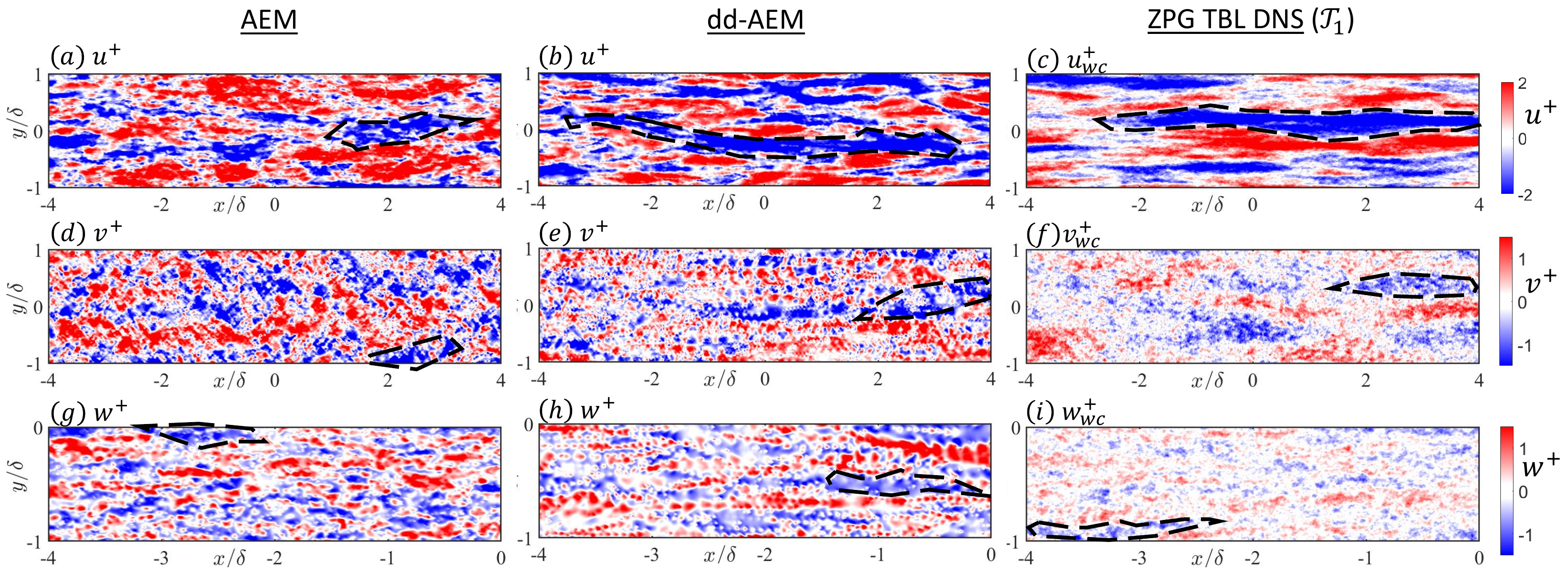}
		\caption{Instantaneous (a,b,c) streamwise, (d,e,f) spanwise and (g,h,i) wall-normal velocity fluctuations on a wall-parallel plane at the lower bound of the inertial-region ($z$ $\approx$ 0.05$\delta$ for the present $Re_{\tau}$). Data in (a,d,g) corresponds to the AEM of \citet{eich2020}, that in (b,e,h) corresponds to the data driven-AEM developed in the present study, while that in (c,f,i) corresponds to the ZPG TBL DNS of \citet{sillero2014} comprising of solely the wall-coherent motions (represented by subscript wc). Black dashed lines are used to highlight the largest spatial features of low momentum for a qualitative comparison. Note the difference in axis limits between (a-f) and (g-i).} 
		\label{fig7} 
	\end{center}
\end{figure*}
\begin{figure*}
	\begin{center}
		\includegraphics[width=1.05\textwidth]{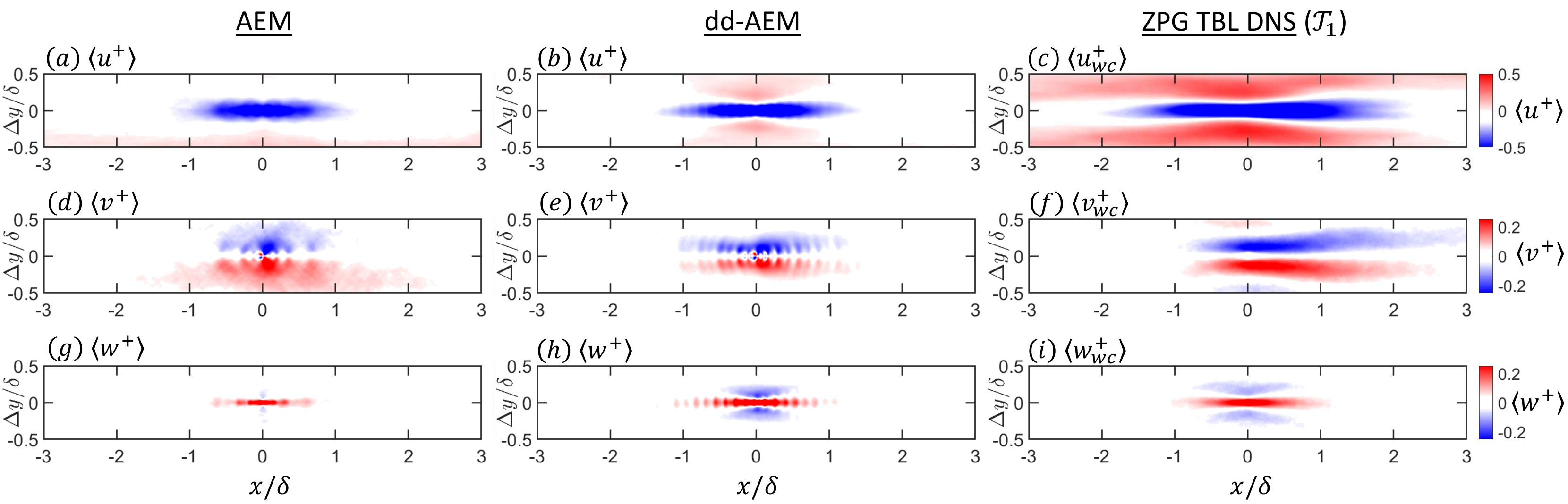}
		\caption{(a,b,c) Streamwise, (d,e,f) spanwise and (g,h,i) wall-normal velocity fluctuations conditioned for ${u}^{+}$ $\lesssim$ -1 and ${w}^{+}$ $\gtrsim$ 1. The conditional average is computed for the flow on a wall-parallel plane at the lower bound of the inertial-region, which is $z$ $\approx$ 0.05$\delta$ for the present $Re_{\tau}$. Data in (a,d,g) corresponds to the AEM of \citet{eich2020}, (b,e,h) corresponds to the dd-AEM developed in the present study, while that in (c,f,i) corresponds to the ZPG TBL DNS of \citet{sillero2014} comprising of solely the wall-coherent motions (represented by subscript $wc$).} 
		\label{fig8} 
	\end{center}
\end{figure*}

\subsection{Spatial representation in the inertial region}

We now extend the comparison between the dd-AEM and the datasets to the instantaneous velocity fluctuations in the inertial region.
Figure \ref{fig7} presents this comparison between the instantaneous fields for all three velocity fluctuations estimated from the dd-AEM, AEM and the ZPG TBL DNS dataset (${\mathcal{T}}_{1}$) at the lower bound of the inertial region.
{Here, it should be noted that the `meandering' aspect of the large-scale motions \citep{hutchins2007} has been built into the dd-AEM framework in the exact same way as in the AEM \citep{eich2020}.}
Amongst these three datasets, the plots associated with the AEM and ZPG TBL DNS are the same as shown previously in figure \ref{fig2}, with significant differences noted between them ($\S$\ref{limit_AEM}).
Now, with the geometry of the representative eddy defined based on data in the dd-AEM, along with the inclusion of the $\delta$-scaled superstructures, the corresponding instantaneous velocity fields (figures \ref{fig7}(b,e,h)) can be observed to closely match the DNS fields.
In particular, the extended coherence of the velocity features ($\gtrsim$ 2$\delta$) noted in all three velocity components in the DNS have been well replicated by the dd-AEM, highlighting the impact of basing the eddy geometry on the data.
This reaffirms the fact that all three components are essentially inter-dependent, and can be modelled via a representative eddy comprising multiple vortex structures (figure \ref{fig1}).
The meandering very-large-scale motions or superstructures which extend beyond 6$\delta$ in length \citep{hutchins2007,lee2011}, as observed in the DNS (figure \ref{fig7}(c)), are also well represented by the dd-AEM (figure \ref{fig7}(b)).
{The dd-AEM, however, appears to overestimate the magnitude of the lateral velocity fluctuations (when compared to the DNS), which is likely an artifact of choosing the simplest shape/arrangement for the representative $\Lambda$-eddy packet in the present study (its optimization is beyond the scope of the present study).}

The qualitative agreement between the dd-AEM and the DNS, showcased in figure \ref{fig7}, can be confirmed quantitatively by generating averaged flow fields conditioned on a statistically dominant feature in the flow.
One such statistical flow feature, which is predominant in the inertial region, is the negative value of the instantaneous momentum flux ($uw$ $<$ 0; \citet{wallace1972}, \citet{deshpande2021uw}), which is carried by coherent motions associated with ejection ($u$ < 0, $w$ > 0) and sweep ($u$ > 0, $w$ < 0) events.
Notably, the velocity field associated with the representative $\Lambda$-eddy packet (figure \ref{fig1}) is also consistent with this notion and models features associated with both ejection and sweep events.
Hence, we compute the conditionally averaged wall-parallel flow fields for the coherent regions associated with strong ejections and sweeps by following the same methodology as adopted previously by \citet{charitha2018} and \citet{eich2020}.
This is performed by attaching a frame of reference to the centroid of each flow feature representing strong $u$ and $w$ fluctuations, which are used as the conditioning points.
Here, the strong fluctuations correspond to $u^+$ $\lesssim$ -1 and $w^+$ $\gtrsim$ 1 for ejections and vice versa for the sweeps.
The mean flow feature is obtained by averaging multiple such strong features extracted from several flow fields.

Figure \ref{fig8} presents the averaged velocity signatures conditioned for the strong ejection features, at the lower bound of the log-region, from flow fields obtained from the AEM, dd-AEM and the DNS.
Firstly, it can be noted that the conditionally averaged fields from the DNS (figures \ref{fig8}(c,f,i)) qualitatively resemble the velocity fields around a single $\Lambda$-eddy packet (figure \ref{fig1}), giving further support to the choice of the representative eddy.
On comparing the geometric extents of the velocity features generated from the two AEMs and the DNS, a better match between the dd-AEM and the DNS can be noted, confirming the enhanced performance of the dd-AEM.
To list a few specific improvements, the long streamwise coherence observed particularly in ${\langle}{w^+_{wc}}{\rangle}$ (figure \ref{fig8}(i)) is well replicated by ${\langle}w^+{\rangle}$ generated from the dd-AEM (figure \ref{fig8}(h)).
Further, the extent of spanwise coherence for all three velocity components, noted in the dd-AEM, compares better with the DNS than that noted for the AEM. 
{The inclusion of relatively long hierarchies in the dd-AEM, which also considers the meandering aspect of the large-scale motions, clearly brings out the distinct X-pattern in the ${\langle}{u^+}{\rangle}$ similar to that seen for ${\langle}{u^+_{wc}}{\rangle}$ from the DNS, and noted previously in the literature \citep{hutchins2007,elsinga2010}.}
The streamwise extent of the wall-parallel velocity features, however, are still falling short of the estimates from the DNS.
This may be an artefact of the assumption of scale-independent yaw angles, imposed onto the $\Lambda$-eddy packets, to replicate `meandering' phenomena in the TBL \citep{eich2020}.
It seems plausible that the scales with the longest streamwise extent (i.e. ${\lambda}_{x}$) `meander' with lesser intensity than those with relatively shorter extent, which may favourably alter the comparison being presented in figure \ref{fig8}.
This, however, remains a subject for future work.

It is also worth noting here that the close match of the conditioned fields from the DNS (figures \ref{fig8}(c,f,i)), with the corresponding flow fields from the $\Lambda$-eddy packet (figure \ref{fig1}), is owing to the consideration of solely the WC motions.
Interested readers may refer to figures \ref{app_fig1}(d,e,f) where full flow fields (WC + WI) have been conditionally averaged based on the same criteria as in figure \ref{fig8}.
The significant mismatch between the velocity features from the WC and full fields, for the DNS, suggests the present representative eddy packet as a suitable choice to model solely the WC subset of the TBL (and not the full flow as a whole).

\section{Discussion}

\begin{figure*}
	\begin{center}
		\includegraphics[width=1.0\textwidth]{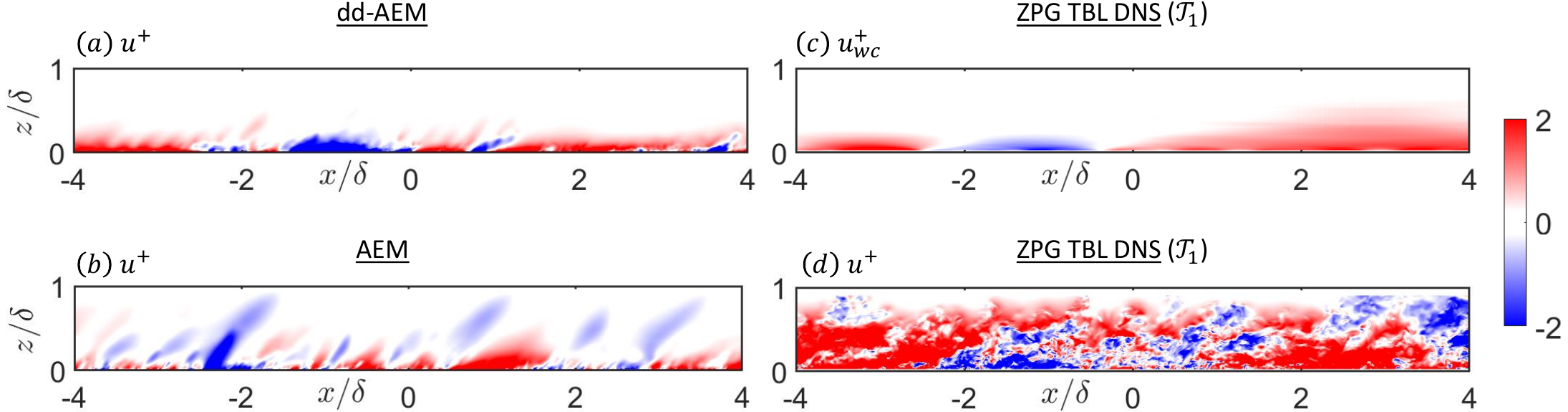}
		\caption{Instantaneous streamwise velocity fluctuations on a wall-normal plane. Data in (a) corresponds to the dd-AEM developed in the present study, that in (b) is from the AEM of \citet{eich2020}, while that in (c,d) corresponds to the same ZPG TBL DNS flow field \citet{sillero2014}, with difference being (c) comprises of solely the wall-coherent motions (represented by subscript wc) and (d) the full flow field (WC+WI).} 
		\label{fig10} 
	\end{center}
\end{figure*}

The present effort promises enhancement in the spatial representation of an instantaneous wall flow by means of a data-driven coherent structure-based model.
One would hope that such models, which have the capability to predict very high $Re_{\tau}$ flows in a computationally inexpensive way, will drive future investigations into flow phenomena otherwise difficult to measure experimentally, such as the instantaneous variation of the $w$-component over the homogeneous wall-parallel plane (as in figure \ref{fig7}(h)) or that of $v$-component over the wall-normal plane.
While the data driven-AEM developed here seems a promising prospect in this regard, more work needs to be done in terms of also incorporating the WI motions to be able to predict the full inertially-dominated turbulent flow field.
These WI motions not only are as statistically significant as the WC motions in the inertial region \citep{chandran2020,charitha2020}, but are also the key driver of the flow phenomena in the wake region.
This is evident from the instantaneous $u$-velocity fluctuations plotted in the wall-normal plane, which are compared for the AEM, dd-AEM and the DNS in figure \ref{fig10}.
As we would expect, the dd-AEM predictions (figure \ref{fig10}(a)) compare well with the instantaneous flow field from the DNS, comprising solely the WC motions (figure \ref{fig10}(c)).
These motions, however, lose their coherence in the wake region where the WI motions predominate, which can be deciphered from the full DNS field in figure \ref{fig10}(d).
{Another aspect of coherent structure-based modelling requiring further work is the shape/form of the representative eddy. 
The present choice of the $\Lambda$-eddy packet for the AEM, although consistent with the statistical picture ($\Gamma$) obtained empirically, does not reflect the typical flow structure observed in an instantaneous flow, leading to overestimation of the instantaneous lateral velocity fluctuations noted in figure 7. 
Efficient machine learning-based algorithms \citep{brunton2020}, which have the capability to analyze big datasets, may be better placed to propose a solution for this.} 

\begin{figure}
	\begin{center}
		\includegraphics[width=0.5\textwidth]{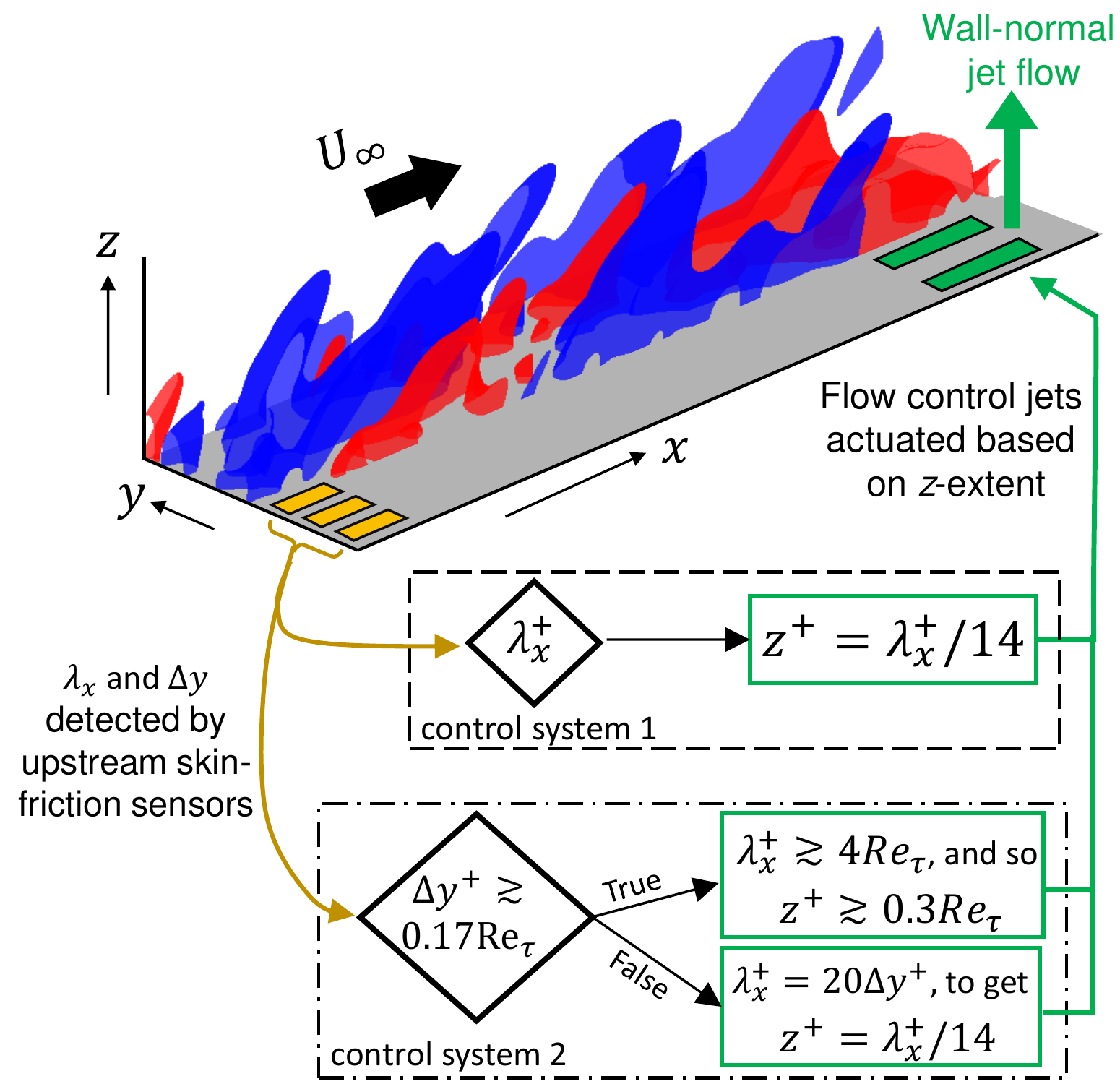}
		\caption{Conceptualization of an active-flow control system aimed at efficiently manipulating the wall-coherent inertial motions in a ZPG TBL by utilizing the scalings determined from high $Re_{\tau}$ datasets. Concept of the flow control system has been adapted from \citet{abbassi2017}. $U_{\infty}$ denotes mean freestream speed.} 
		\label{fig9} 
	\end{center}
\end{figure}

Besides assisting with the modelling of the WC motions, the empirically determined geometric scalings in (\ref{eq2},\ref{eq3}) can also be used to enhance the active flow control schemes operating based on real-time sensing.
Figure \ref{fig9} depicts a conceptual sketch showcasing the coherent motions in a ZPG TBL, which can be manipulated by cross-flow (wall-normal) jets controlled by a computer, based on data from a spanwise array of skin-friction sensors placed at a sufficiently upstream location \citep{abbassi2017}.
These skin-friction sensors, which are used to detect the incoming high-momentum ($+u$) carrying coherent motions, are also capable of detecting the streamwise extent (${\lambda}_{x}$) of these motions.
Utilizing the empirically estimated scalings, the computer controlling the jet actuation can decide how much momentum to inject through them, based on the estimated wall-normal extent ($z$) of the incoming motions (indicated as control system 1 in figure \ref{fig9}).
Alternatively, in scenarios where sufficiently long separations between the sensors and the jets cannot be maintained, the spanwise sensor array can be used to estimate the spanwise coherence (${\Delta}y$) of the incoming motions, through which the wall-normal extent could be predicted (indicated as control system 2 in figure \ref{fig9}). 
Such empirically-driven flow control systems, which also consider the wall-normal extent of the incoming motions, could potentially manipulate the turbulent inertial wall flow more efficiently, as demonstrated previously by \citet{yao2017}.
Given the significant contribution of the attached eddy hierarchy to the mean skin friction at high $Re_{\tau}$ \citep{giovanetti2016}, availability of their corresponding 3-D geometric scalings can also be used to decide the spanwise spacing between the cross-flow jets targeting these eddies.

\section{Concluding remarks}

The present study analyses a unique set of multi-point datasets to reconstruct the 3-D statistical picture of the inertial wall-coherent (WC) turbulence in all three canonical wall-bounded flows.
Previous studies have found these motions to be responsible for both, the $Re_{\tau}$-dependence of the skin-friction drag \citep{orlu2011,deck2014,giovanetti2016,smits2021} as well as the bulk production and the inter-scale energy transfer in high $Re_{\tau}$ flows \citep{marusic2010high,mklee2019,yhwang2020}. 
The aforementioned characteristics make these motions a key target of coherent structure-based models, which the present study attempts to enhance.

Here, the statistical picture is reconstructed by computing the cross-correlation spectra ($\Gamma$) using the streamwise velocity fluctuations mapped across the 3-D space in the wall-bounded shear flow.
The intermediate- and large-scaled inertial WC motions are found to exhibit geometric self-similarity with respect to the distance from the wall $z$, expressible by simple linear relationships, which are universal across all canonical flows and independent of flow $Re_{\tau}$.
The geometry of the very-large-scaled motions, on the other hand, is found to exhibit $\delta$-scaling, associating them with the superstructures and VLSMs for external and internal flows, respectively.
The present study also confirms the 3-D geometry of the VLSMs to be much larger than the superstructures even in high $Re_{\tau}$ flows, highlighting the role of the T/NTI in inhibiting the growth of these very-large-scaled motions in external flows \citep{monty2007,lee2013}.

Alongside testing the universality of the 3-D statistical picture, the geometric scalings brought out from the data (given by \ref{eq2} and \ref{eq3}) are also proposed to be used as a metric to estimate the spanwise and wall-normal extent corresponding to each WC scale/eddy (${\lambda}_{x}$).
Application of these linear relationships to any coherent structure-based model can provide a data-driven basis to the geometry of the representative structure, which is demonstrated in this study using the attached eddy model (AEM; \citet{perry1982}) for a ZPG TBL.
Here, the choice of AEM (over other models) is also driven by the empirical analysis, given the latter provides evidence of geometric self-similarity of the inertial motions and also supports the $\Lambda$-eddy packet as the representative eddy (both of which are built into the AEM framework; \citet{marusic2019}).
The present study further extends the scope of the AEM by using its framework to also model the $\delta$-scaled superstructures, which correspond to the very-large-scales beyond the self-similar hierarchy.
The data-driven AEM (or dd-AEM), which has the representative eddies defined based on the data, is shown to improve upon recent works on the AEM \citep{eich2020} in replicating instantaneous flow phenomena associated with all three velocity components.
This sets up the platform for future work, which would include both the WC and WI (wall-incoherent) motions modelled via the AEM framework, to replicate the full turbulent wall-bounded flow that can provide realistic inflow conditions for use in future numerical simulations \citep{subbareddy2006,wu2017}.

\section*{Acknowledgement}
The authors wish to acknowledge the Australian Research Council for financial support. 
They are also grateful to the authors of \citet{sillero2014}, \citet{baars2017} and \citet{baidya2019} for making their respective data available.
Sandia National Laboratories is a multimission laboratory managed and operated by National Technology and Engineering Solutions of Sandia, LLC., a wholly owned subsidiary of Honeywell International, Inc., for the U.S. Department of Energy's National Nuclear Security Administration under contract DE-NA0003525. 
This paper describes objective technical results and analysis. 
Any subjective views or opinions that might be expressed in the paper do not necessarily represent the views of the U.S. Department of Energy or the United States Government.

\appendix
\section{SLSE-based decomposition of the instantaneous flow fields}
\label{appendix}

\begin{figure*}
	\begin{center}
		\includegraphics[width=1.0\textwidth]{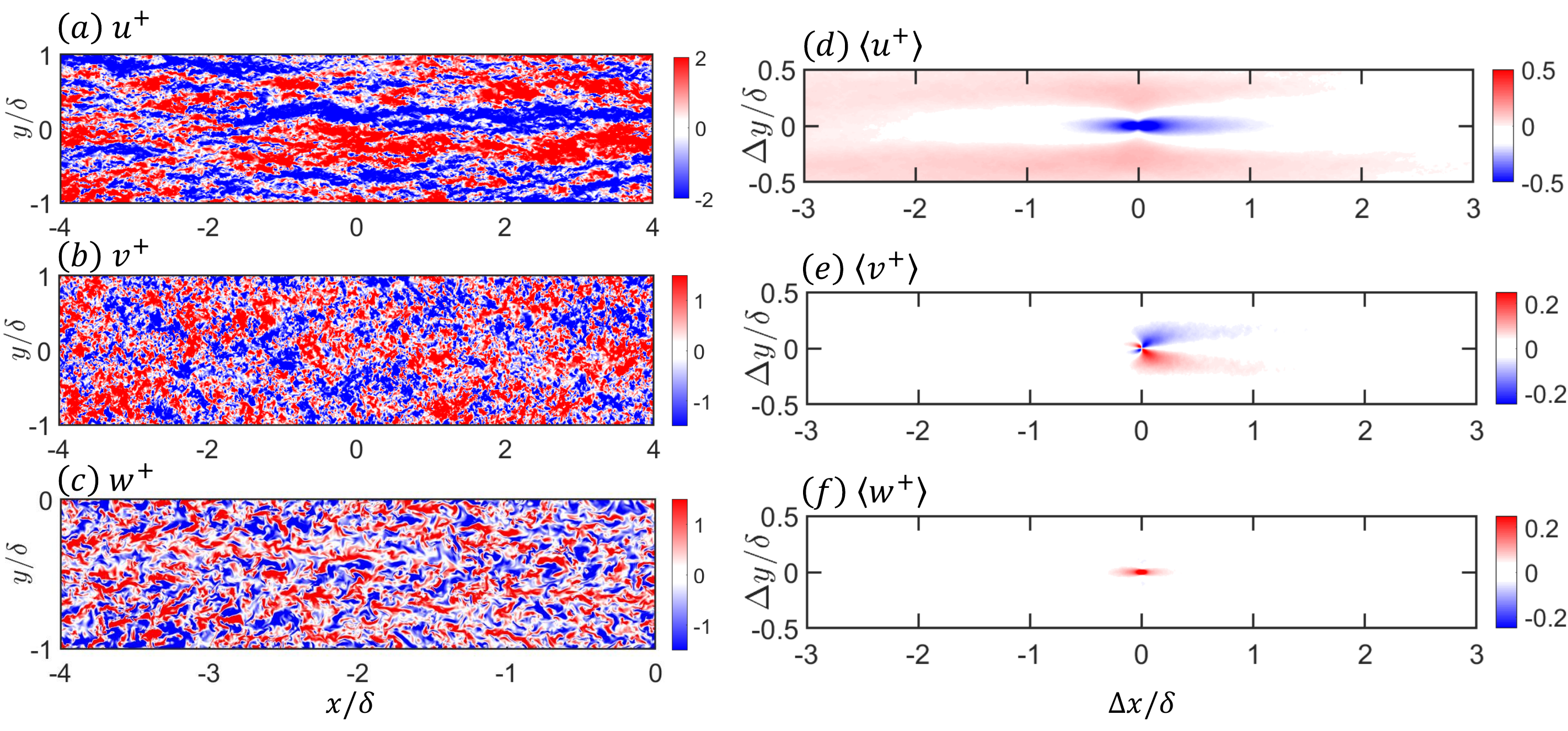}
		\caption{Instantaneous (a) streamwise, (b) spanwise and (c) wall-normal velocity fluctuations on a wall-parallel plane at the lower bound of the log-region ($z$ $\approx$ 0.05$\delta$ for the present $Re_{\tau}$) extracted from ZPG TBL DNS dataset (${\mathcal{T}}_{1}$) of \citet{sillero2014}. Unlike figures \ref{fig2} or \ref{fig7}, these flow fields comprise contributions from both WC and WI motions, i.e. they are the raw flow fields directly from the dataset. (d) Streamwise, (e) spanwise and (f) wall-normal velocity fluctuations conditioned for $u^+$ $\lesssim$ -1 and $w^+$ $\gtrsim$ 1 on the same wall-parallel plane as in (a-c).} 
		\label{app_fig1} 
	\end{center}
\end{figure*}
 
As discussed previously in $\S$\ref{intro}, the flow field in the inertial region comprises contributions from both wall-coherent (WC) and wall-incoherent (WI) eddies.
Here, we demonstrate a methodology, based on the spectral linear stochastic estimate (SLSE) approach, to estimate the WC subset of a velocity fluctuation (say for $u$-component) at any wall-normal location $z$ in the inertial region.
This procedure has been utilized in previous studies \citep{baars2016slse,anagha2019,deshpande2021uw} for similar purposes (i.e. to obtain a subset of the full flow field comprising selected coherent motions), which can be directly referred for an elaborate introduction to this technique.
According to SLSE, a scale-specific unconditional input at any near-wall location, $z_{w}$ (with $z^+_w$ $\lesssim$ 15) can be used to obtain a scale-specific conditional input at $z$ following:
\begin{equation}
\label{app_eq1}
%\begin{split}
{\widetilde{u}}^{E}(z^{+};{\lambda}^{+}_{x},{\lambda}^{+}_{y}) = {{H^{u}_{L}}(z^{+},z^{+}_{w};{\lambda}^{+}_{x},{\lambda}^{+}_{y})}\widetilde{u}(z^{+}_{w};{\lambda}^{+}_{x},{\lambda}^{+}_{y}), 
%\end{split}
\end{equation}
where ${\tilde{u}}({z_{w}};{\lambda}_{x},{\lambda}_{y})$ is the 2-D Fourier transform of the instantaneous wall-parallel flow field, $u$(${z_w};x,y$) in space.
Here, the superscript $E$ and $H^{u}_{L}$ respectively represent the estimated quantity and the scale-specific linear transfer kernel (for $u$-component).
$u^{E}$ in equation (\ref{app_eq1}), essentially, corresponds to the energy distribution (at $z$) across wavelengths ${\lambda}_{x}$ and ${\lambda}_{y}$, which are coherent across $z$ and $z_w$.
This is facilitated by first computing $H^{u}_{L}$ from an ensemble of data as per:
\begin{equation}
\label{app_eq2}
%\begin{aligned}
{H^{u}_{L}}(z^{+},z^{+}_{w};{\lambda}^{+}_{x},{\lambda}^{+}_{y}) = \frac{ \{ \widetilde{u}(z^{+};{\lambda}^{+}_{x},{\lambda}^{+}_{y}){{\widetilde{u}}^{\ast}}(z^{+}_{w};{\lambda}^{+}_{x},{\lambda}^{+}_{y}) \} }{ \{ {\widetilde{u}(z^{+}_{w};{\lambda}^{+}_{x},{\lambda}^{+}_{y})}{{\widetilde{u}}^{\ast}(z^{+}_{w};{\lambda}^{+}_{x},{\lambda}^{+}_{y})} \} },
%\end{aligned}
\end{equation}
where the curly brackets ($\{ \}$) and asterisk ($\ast$) denote the ensemble averaging and complex conjugate, respectively.
Given that $z^+_{w}$ $\lesssim$ 15 and $z^+_{w}$ $\ll$ $z^+$, ${\widetilde{u}}^{E}(z^{+};{\lambda}^{+}_{x},{\lambda}^{+}_{y})$ thus represents energy contributions at $z$ from solely WC motions, that are taller than $z$ \citep{anagha2019,deshpande2021uw}, leading to:
\begin{equation}
\label{app_eq3}
{{{{\widetilde{u}}^{E}}(z^{+};{\lambda}^{+}_{x},{\lambda}^{+}_{y})}{{\big|}_{{z^{+}_{w}} {\lesssim} 15}}}\; {\rightarrow}\; {{{\widetilde{u}}_{wc}}(z^{+};{\lambda}^{+}_{x},{\lambda}^{+}_{y})},
\end{equation}
where $u_{wc}$($z$) represents the subset of $u$($z$) comprising solely the WC motions.
On obtaining ${{\widetilde{u}}_{wc}}$, the corresponding flow field in physical space can be obtained by:
\begin{equation}
\label{app_eq4}
{{{u}_{wc}}(z;x,y)} = {\mathcal{F}^{-1}}({{{\widetilde{u}}_{wc}}(z;{\lambda}_{x},{\lambda}_{y})}),
\end{equation}
where ${\mathcal{F}^{-1}}$ represents the inverse Fourier transform.
Here, although the entire procedure has been demonstrated solely for $u$-fluctuations, the same can be applied to other (lateral) velocity fluctuations by estimating the respective linear transfer kernel ($H^{v}_{L}$ or $H^{w}_{L}$).
To give an example, figures \ref{app_fig1}(a-c) plot the instantaneous wall-parallel flow fields, comprising both WC + WI motions (i.e. the full flow fields) at the lower bound of the inertial region.
While, figures \ref{fig2}(b,d,f) plot the subset of the corresponding full fields comprising solely the WC motions, estimated using the aforementioned procedure.

\bibliographystyle{cas-model2-names}

\bibliography{Deshpande_DataDrivenAEM2021_FinalArxivSubmission_bib}

\end{document}